\documentclass[aps,prb,reprint]{revtex4-1}

\usepackage{amsmath,euscript,theorem,graphicx,rotating}
\usepackage[latin1]{inputenc}
\usepackage[T1]{fontenc}
\usepackage{txfonts}
\usepackage{url}
\usepackage{xr} 
\usepackage{color}
\usepackage{dcolumn}
\usepackage{bm}
\usepackage{longtable}
\usepackage{xspace}
\usepackage{booktabs}
\usepackage{version}

\usepackage{array}

\newcommand{\abinitio}{{\em ab initio}\xspace}

\newcommand{\etal}{\emph{et al.}\xspace}
\newcommand{\WSM}{WSM\xspace}

\newcommand{\dr}{{\mathrm d}^{3}{\mathbf r}}
\newcommand{\drp}{{\mathrm d}^{3}{\mathbf r'}}
\newcommand{\rr}{{\mathbf r}}

\newcommand{\RR}{{\mathbf R}}
\newcommand{\rp}{{\mathbf r'}}

\newcommand{\secrff}[1]{{\ \S \ref{#1}}}

\newcommand{\figrff}[1]{Figure\ \ref{#1}}
\newcommand{\tabrff}[1]{Table\ \ref{#1}}

\newcommand{\eqrff}[1]{eq.~\eqref{#1}}

\newcommand{\order}[1]{\ensuremath{\mathcal{O}(#1)}\xspace}
\newcommand{\neighbourhood}[1]{\ensuremath{\mathcal{N}_{#1}}\xspace}
\newcommand{\fdds}{\ensuremath{\alpha(\rr,\rp|\omega)}\xspace}
\newcommand{\fddsdist}[1]{\alpha^{#1}(\rr,\rp|\omega)}

\newcommand{\pol}[2]{\alpha^{#1}_{#2}(\omega)}

\newcommand{\Q}[2]{\ensuremath{Q^{#1}_{#2}}\xspace}
\newcommand{\oQ}[2]{\ensuremath{\hat{Q}^{#1}_{#2}}\xspace}

\newcommand{\C}[2]{\ensuremath{C^{#1}_{#2}}\xspace}
\newcommand{\Cn}[1]{\ensuremath{C_{#1}}\xspace}

\newcommand{\bb}[1]{\ensuremath{\beta_{#1}}\xspace}
\newcommand{\tbb}[1]{\ensuremath{\tilde{\beta}_{#1}}\xspace}

\newcommand{\EDISP}[1]{\ensuremath{E^{(#1)}_{\rm DISP}}\xspace}
\newcommand{\Edisppol}[1]{\ensuremath{E^{(#1)}_{\rm disp,pol}}\xspace}

\newcommand{\Edispexch}[1]{\ensuremath{E^{(#1)}_{\rm disp,exch}}\xspace}

\newcommand{\CamCASP}[1]{{\sc CamCASP #1}\xspace}

\newcommand{\SAPT}{{\sc Sapt2008}\xspace}

\newcommand{\DALTON}{{\sc DALTON} 2.0\xspace}
\newcommand{\Dalton}{\DALTON}
\newcommand{\NWChem}{{\sc NWChem} 6.x\xspace}
\newcommand{\PsiF}{{\sc Psi4} 1.1\xspace}
\newcommand{\Gamess}{{\sc GAMESS(US)} \xspace}

\newcommand{\MCpBS}{MC\ensuremath{^+}\xspace}

\newcommand{\DCpBS}{DC\ensuremath{^+}\xspace}

\newcommand{\w}[1]{\ensuremath{w^{#1}}\xspace}
\newcommand{\R}[1]{\ensuremath{\rho^{#1}}\xspace}
\newcommand{\wL}[1]{\ensuremath{w_{\rm L}^{#1}}\xspace}
\newcommand{\wT}[1]{\ensuremath{\tilde{w}^{#1}}\xspace}

\newcommand{\p}[1]{\ensuremath{p^{#1}}\xspace}
\newcommand{\ISAfunc}[1]{\ensuremath{\Delta_{\rm stock(#1)}}\xspace}

\newcommand{\cDF}[0]{cDF\xspace}
\newcommand{\cDFL}[0]{cDF-L\xspace}

\newcommand{\SRLO}[0]{SRLO\xspace}
\newcommand{\SRLOL}[0]{SRLO-L\xspace}

\newcommand{\ISApol}[0]{ISA-Pol\xspace}

\newcommand{\ISApolL}[0]{ISA-Pol-L\xspace}

\mathchardef\lt="313C \mathchardef\gt="313E
\mathcode`<="4268 \mathcode`>="5269
\newcolumntype{d}[1]{D{.}{.}{#1}}
\newcolumntype{.}{D{.}{.}{-1}}
\newcolumntype{,}{D{,}{,}{2}}

\newcolumntype{L}{>{$}l<{$}}
\newcolumntype{R}{>{$}r<{$}}
\newcolumntype{C}{>{$}c<{$}}

\newcommand{\vdw}{\mbox{\ensuremath{\cdot\!\cdot\!\cdot}}}

\newcommand{\JCP}[0]{J. Chem. Phys.\ }
\newcommand{\JPCA}[0]{J. Phys. Chem. A\ }

\newcommand{\JCTC}[0]{J. Chem. Theory Comput.\ }

\newcommand{\CPL}[0]{Chem. Phys. Lett.\ }
\newcommand{\TCA}[0]{Theor. Chim. Acta\ }

\newcommand{\PRA}[0]{Phys. Rev. A\ }
\newcommand{\PRB}[0]{Phys. Rev. B\ }

\newcommand{\PRL}[0]{Phys. Rev. Lett.\ }

\newcommand{\MolP}[0]{Mol. Phys.\ }

\newcommand{\PCCP}[0]{Phys. Chem. Chem. Phys.\ }
\newcommand{\JCC}[0]{J. Comp. Chem.\ }
\newcommand{\JACS}[0]{J. Am. Chem. Soc.\ }

\newcommand{\IRPC}[0]{Int. Revs. Phys. Chem.\ }

\newcommand{\ChemComm}[0]{Chem. Commun.\ }

\begin{document}

\title[Distributed polarizabilities from the ISA]
{\ISApol: Distributed polarizabilities and dispersion models from a
basis-space implementation of the iterated stockholder atoms procedure}

\author{Alston J. Misquitta}
\affiliation{School of Physics and Astronomy, 
and the Thomas Young Centre for Theory and Simulation of Materials,
Queen Mary, University of London,
London E1 4NS, U.K.}
\author{Anthony J. Stone}
\affiliation{University Chemical Laboratory, Lensfield Road,
Cambridge, CB2 1EW, U.K.}

\date{\today}

\begin{abstract}
Recently we have developed a robust, basis-space implementation of the
iterated stockholder atoms (BS-ISA) approach for defining atoms in a 
molecule.
This approach has been shown to yield rapidly convergent distributed multipole 
expansions with a well-defined basis-set limit. 
Here we use this method as the basis of a new approach, termed \ISApol,
for obtaining non-local distributed frequency-dependent polarizabilities.
We demonstrate how \ISApol can be combined with localization methods 
to obtain distributed dispersion models that share the many unique properties
of the ISA: These models have a well-defined basis-set limit, lead to very accurate 
dispersion energies, and, remarkably, satisfy commonly used combination
rules to a good accuracy. 
As these models are based on the ISA, they can be expected to respond to
chemical and physical changes naturally, and thus they may serve as the basis
for the next generation of polarization and dispersion models
for \abinitio force-field development.
\end{abstract}

\maketitle



\section{Introduction}
\label{sec:introduction}
In the last few years, the field of intermolecular interactions has seen a tangible
increased level of importance. The deep level of understanding we have 
achieved from decades of theoretical developments has formed the basis of 
new models for intermolecular interactions that finally give us the promise 
of the long-awaited accuracy and predictive power needed in application to 
complex molecular aggregation processes. 

These intermolecular interaction models are being developed primarily 
from interaction energies computed using some variant of 
symmetry-adapted perturbation theory (SAPT), and predominantly using
the version of SAPT based on density-functional theory, SAPT(DFT). 
The latter choice is based both on the favourable accuracy and
computational efficiency of SAPT(DFT). 
The general procedure for model development typically uses some mix
of SAPT(DFT) calculations at specific, close-separation dimer configurations,
and an analytical multipole-expanded form of the 
interaction energy suitable for the long range.
The various implementations of this approach have been described elsewhere
\cite{MisquittaS16,VanVleetMSS15,MetzPS16,VandenbrandeWSV17,VanVleetMS18}.

The advantage of using a theory like SAPT or SAPT(DFT) for the short-range
energies is that the resulting interaction energy has a well-defined
multipole-expanded form. Consequently, if this multipole-expanded form
can be determined analytically, there can be a rigorous match between 
the short and long range. 
Indeed, this has been the basis of the above philosophy for many decades
(see for example refs.~\cite{MasSBJ97,KoronaWBJS97,BukowskiSC99,ChangA-OBS03,MisquittaBS00,MurdachaewMBS01}).
Here SAPT(DFT) has an advantage over SAPT in that the multipolar molecular
properties (multipole moments, polarizabilities, dispersion coefficients)
can be readily derived from the underlying density functional method,
and usually at a comparatively low computational cost.

However, as is now well known
\cite{StoneM07,Stone:book:13,RobS13a,GagliardiLK04,HarczukNJVA17,MisquittaS06,MisquittaS08a,MisquittaSP08,AngyanJLHH94,DehezCMA01,JansenHHA96,AngyanCDHJM03,LillestolenW07,WheatleyL08,WheatleyM94,ChaudretEtAl14,VigneC88}
intermolecular properties must be {\em distributed} if we are to achieve 
high enough accuracies. The single-centre multipole expansion, which is
a useful paradigm for diatomics or triatomics, is poorly convergent 
for larger molecules, for which we must use multiple expansion centres.
These expansion centres have usually been taken to be the locations of the 
nuclei in the molecule, though this need not be the case, and indeed, 
for some cases \cite{MasBSG00,FangDRM13} multiple, off-atomic sites are chosen to 
obtain even faster convergence of the multipole expansion.

The problem with calculating distributed properties is that it does not
seem possible to define a unique way of partitioning a molecular 
property into portions associated with the atoms in a molecule (AIMs).
This ambiguity has led to a whole range of schemes to define the AIMs
(see for example Refs.~\citenum{Bader90,Hirshfeld77,LillestolenW08,BultinckCN09}), 
which have, in turn, resulted in some lively discussion in the published
literature \cite{ParrAN05,MattaB06}.
Here we do not wish to address the more philosophical issues associated
with the atom-in-a-molecule, but rather focus on some of the practicalities
that result from the choice of AIM method. 
Consider the following list of features of the distributed molecular properties
that we might like to see achieved:
\begin{itemize}
  \item {\em Uniqueness for a given choice of AIM algorithm}: 
    While the AIMs themselves are not unique, the actual atomic domains
    that result from a particular choice of partitioning algorithm should
    be unique. That is, the result should not depend on numerical parameters,
    and should have a well defined basis-set limit. This will usually imply that
    the resulting distributed molecular properties are also unique. 
  \item {\em Rapid convergence with rank}: As the distributed properties
    will typically be used in a model for the molecular interactions, for 
    computational reasons it is usually desirable that these models 
    be rapidly convergent with rank.
    This condition implies that the atomic domains from the AIM are
    as close to being spherical as is possible.
  \item {\em Agreement with reference energies}:
    The distributed properties should result in energies in good agreement
    with those from the reference electronic structure method. 
    In our case this will be taken to be appropriate interaction
    energies from SAPT(DFT).
  \item {\em Insensitivity to molecular conformation}: 
    We fully expect distributed properties to vary with molecular
    conformation, but, particularly for {\em soft} deformations, that is
    those with a small change in the electronic distribution, we
    may expect the AIM domains and resulting molecular properties
    also to change only slightly.
  \item {\em Agreement with physical/chemical expectations}:
    This condition is qualitative as we cannot define what the physically
    meaningful properties of an atom in a molecule should be. 
    We can however hope that the resulting properties be in broad agreement
    with chemical/physical intuition.
  \item {\em Computational efficiency}: This is important if we are
    to apply the distribution techniques to large systems.
    Ideally we would like the algorithm to scale linearly with the 
    size of the system.
\end{itemize}
Not all of these requirements need to be met to develop an interaction 
model for a specific system: after all, the long-range parameters 
can be treated as fitting parameters chosen to result in the best 
fit to the reference energies.
However the parameters resulting from such a mathematical fit rarely 
have any link to the physical properties of the system, and consequently 
cannot be used for the development of more general interaction models.
Instead we must turn to methods that are somehow linked to the underlying
properties of the atom in a molecule.

Some of the methods used to define the properties of the atoms in a molecule
can be regarded as being more mathematical or numerical, though physical
properties like the van der Waals radii may be used.
In these methods, the molecular properties may be partitioned
in a basis-space or real-space manner, though hybrids of the two are
also used. Some of the more successful of these methods include the
distributed multipole analysis (DMA) of Stone \cite{AJSxxxvii,Stone05},
the LoProp and MpProp approaches \cite{GagliardiLK04,SoderhjelmKKRL07},
and methods based on constrained density fitting for the 
multipole moments \cite{PiquemalCRG06} and for the 
polarizabilities \cite{MisquittaS06,RobS13a}. 
We will refer to the original constrained density-fitting method of
Misquitta \& Stone \cite{MisquittaS06} as the \cDF method, and the
related `self-repulsion plus local orthogonality' method of 
Rob \& Szalewicz \cite{RobS13a} as the \SRLO method. 

Both the \cDF and \SRLO distribution techniques use constraints in the
density fitting to allow the molecular polarizabilities to be partitioned
into non-local, site-site polarizabilities. These are not the local
polarizabilities that one might conventionally think of, but include terms that
allow for non-local, or through-space polarization in the
molecule.\cite[\S9.2]{Stone:book:13} 
The methods differ in the constraints applied, with the \SRLO algorithm
using a constraint to reduce the charge-flow terms, that is, the polarizabilities
that allow for charge movement in the molecule, to nearly zero. 
Using appropriate localization techniques \cite{MisquittaS08a,MisquittaSP08}
both the \cDF and \SRLO models can be made to yield effective 
local polarizability models. In the case of the former, we have referred
to the combined method as the Williams--Stone--Misquitta, or \WSM model.
This model has formed the basis of much of our work so far, and indeed
has been used to develop intermolecular interaction models by 
other groups either directly \cite{SebetciB10a} or by extension 
\cite{McDanielS13,VanVleetMSS15,VanVleetMS18}.
As the localization schemes in the \WSM model can be applied to any of
the non-local polarizability models, we will refer to the localized
models by appending `-L', for example the \SRLOL model would be the 
\SRLO non-local model localized using the \WSM approach.

While these methods have been successful in developing useful models
for both the polarization and the dispersion energies, the AIM properties
resulting from either the \SRLOL or \cDFL algorithms do not have a 
well-defined basis-set limit and can result in unexpected, and perhaps
unphysical AIM properties. Consider the \cDFL localized, {\em isotropic}
polarizabilities for the thiophene molecule shown in 
\tabrff{table:thiophene-cDF-pols}.
While the dipole-dipole polarizabilities for all sites appear to be 
reasonably stable with basis with variations of 5\% or so, the
same cannot be said for the higher ranking polarizabilities:
there are significant variations with basis set in the 
quadrupole-quadrupole polarizabilities, 
with negative values for the two hydrogen AIMs in the 
triple-$\zeta$ basis, and the octopole-octopole AIM polarizabilities are
negative for most of the data in the table. 
We note that even though these individual polarizabilities appear 
unphysical, the whole description yields the correct total molecular
polarizability.
The \SRLOL polarizability models yield much the same picture and are not
shown.
These problems can be partially reduced by constraining the localization 
or by including more data during the refinement steps of the \WSM method
as indeed has been done by McDaniel and Schmidt \cite{McDanielS13}, 
but an alternative is needed.

\begin{table}[ht]
   \newcolumntype{d}{D{.}{.}{2}}
   \begin{tabular}{llddd}
     \toprule
       Site  & $l$ &\multicolumn{1}{r}{\text{aDZ}}
                               &\multicolumn{1}{r}{\text{aTZ}}
                                         &\multicolumn{1}{r}{\text{aQZ}}\\
     \midrule
       C1    &  1  &     7.28  &   7.20  &    6.94  \\
             &  2  &    28.52  &  32.77  &   21.33  \\
             &  3  &  -355.10  & 141.68  &  920.37  \\
     \midrule
       C2    &  1  &    10.60  &  10.79  &   11.19  \\
             &  2  &    36.47  &  57.05  &   44.03  \\
             &  3  &  -345.20  &-341.59  &  580.76  \\
     \midrule
       S     &  1  &    16.67  &  16.90  &   16.86  \\
             &  2  &    90.78  &  95.12  &  107.16  \\
             &  3  &  -206.73  &-617.48  & -315.64  \\
     \midrule
       H1    &  1  &     2.24  &   2.26  &    2.36  \\
             &  2  &     1.51  &  -3.42  &    6.14  \\
             &  3  &   -69.59  & -38.73  & -155.99  \\
     \midrule
       H2    &  1  &     1.55  &   1.48  &    1.33  \\
             &  2  &     2.88  &  -3.66  &   10.09  \\
             &  3  &   -49.55  & -45.20  & -157.65  \\
     \bottomrule
   \end{tabular}
   \caption{
     Localized, isotropic polarizabilities for the symmetry-distinct
     sites in the thiophene molecule computed with the \cDFL model, that
     is using \cDF non-local polarizabilities localized using the
     \WSM algorithm.
     The basis sets used are aug-cc-pVDZ (aDZ), aug-cc-pVTZ (aTZ),
     and aug-cc-pVQZ (aQZ). 
     Atom C1 is the carbon atom attached to the sulfur atom and H1 is the 
     hydrogen atom attached to C1. 
     Atomic units used for all polarizabilities.
   }
   \label{table:thiophene-cDF-pols}
\end{table}

Consider the more physically motivated schemes to define the AIMs.
These include Bader's topological analysis 
(the so-called quantum theory of atoms-in-a-molecule, or QTAIM)
\cite{Bader90}, maximum probability domain (MPD) analysis
\cite{MenendezMBS15}, and the various methods based on the Hirshfeld 
stockholder partitioning \cite{Hirshfeld77,LillestolenW08,BultinckCN09}.
The method of Bader is perhaps the most well known of the AIM techniques
and has been used for defining both distributed multipole moments 
and polarizabilities \cite{JansenHHA96,AngyanCDHJM03,KosovP00a,PopelierR03}
and has also been used to construct distributed dispersion models
\cite{HattigJHA97}.
However while this technique satisfies a number of the properties 
listed above, it results in unusual AIM domains that lead to a somewhat
slower convergence with rank of the expansion.
The MPD approach is relatively new and has not yet been used as a means
of obtaining distributed properties, but like the 
QTAIM method it is well defined.
The Hirshfeld-like methods are appealing in their simplicity: 
If we define reference, usually spherically-symmetrical
atomic densities $\w{a}(\rr)$ for atom $a$
--- we shall term these the {\em shape functions} (though in other
papers \cite{Ayers06a} this term is used for these functions normalized to unity) --- 
then the density allocated to atom $a$ in the molecule with total
electronic density $\rho(\rr)$ is given by 
\begin{align}
  \R{a}(\rr) &= \rho(\rr) \times \frac{\w{a}(\rr)}{\sum_b \w{b}(\rr)}.
  \label{eq:stockholder-partitioning}
\end{align}
Notice that even if the shape functions are spherically symmetrical, the
AIM density $\R{a}$ will normally be anisotropic.
This scheme for partitioning the molecular density is not only elegant,
but results in smooth, nearly spherical AIM densities
which satisfy many of the requirements we have listed above. 
However there are problems with the original Hirshfeld scheme in which the
reference atomic densities were chosen to be 
the densities of the isolated, neutral atoms. 
This has been recognised \cite{BultinckVAAC-D07,BultinckCN09} to be a poor choice as it
causes the AIM densities to be as similar as possible to the neutral free atoms
with the consequence that charge movement in the molecule was sometimes
severely underestimated. 
Bultinck \etal \cite{BultinckVAAC-D07,BultinckCN09} provided an elegant solution to this
problem by allowing the reference state to be a linear combination of
free ionic states, with the occupancy probabilities being determined 
self-consistently in what is known as the Hirshfeld-I scheme.

An even more elegant solution to the problem of the original Hirshfeld scheme
was proposed by Lillestolen \& Wheatley \cite{LillestolenW08} who 
proposed that the reference atomic densities be determined 
{\em self-consistently} by defining them as the spherical average of the
AIM densities:
\begin{align}
  \w{a}(\rr) &= < \R{a}(\rr) >_{\rm sph}.
  \label{eq:ISA-self-consistency}
\end{align}
This method, termed the iterated stockholder atoms (ISA) algorithm 
requires no {\em a priori} reference states. Instead, once a guess to the
states is made, \eqrff{eq:stockholder-partitioning} and \eqrff{eq:ISA-self-consistency}
are iterated to self-consistency to achieve the desired solution.
Early attempts at finding the ISA solution often needed as many as a 
thousand iterations to reach convergence, and sometimes failed to
converge at all, but more robust algorithms have recently been
developed that generally achieve convergence in a few dozen
iterations.\cite{VerstraelenAvanSW12,MisquittaSF14}
These new methods work by restricting the variational freedom given to the
ISA reference functions by defining them via a basis expansion rather than
in real space as was formerly done.

One of these methods is the basis-space ISA, or BS-ISA algorithm
that we have developed and implemented in the \CamCASP{} \cite{CamCASP} 
program. We have used the BS-ISA algorithm to define distributed
multipole models and have demonstrated that these multipoles 
exhibit all of the properties we have listed above.
In fact, the BS-ISA distributed multipoles --- or ISA-DMA models for short ---
surpass those from the well-established distributed multipole analysis (DMA)
algorithm by Stone \cite{Stone85,Stone05} in the rapidity of convergence
with rank and in the stability with respect to basis set. 
Further, we have demonstrated how the BS-ISA density partitioning
can be used, via the distributed overlap model, to achieve robust 
fits to the short-range part of the interaction energy and thereby 
to easily develop detailed analytic models for the 
intermolecular interaction \cite{MisquittaS16}. 
Finally, in collaboration with Van Vleet and Schmidt \cite{VanVleetMSS15,VanVleetMS18}
data from the BS-ISA algorithm has been used to develop the short-range repulsion and
dispersion damping models for two general force fields: 
the Slater-FF and MASTIFF models.

In this paper we extend the applicability of the BS-ISA algorithm to the
second-order energies and we demonstrate how we can use this method to 
obtain distributed frequency-dependent polarization models, and from these,
distributed dispersion models for any closed-shell molecular system.
We first describe this new algorithm, termed \ISApol. 
Next we describe a new, simplified and more flexible version of the BS-ISA
algorithm, one that allows more accurate ISA solutions as well as additional 
sites and coarse-graining. 
The \ISApol method results in what are known as {\em non-local}
polarizabilities which describe through-space polarization
and charge movement in the system. While this is an important subject and
leads to unexpected van der Waals interactions \cite{MisquittaSSA10,LiuAD11,MisquittaMDSN14}
in low-dimensional systems, we will instead focus here on the 
{\em localized} distributed models that lead to the conventional 
polarization and dispersion interactions. 
We describe the localization procedures in brief along with some of the
important features of the methods.
Then we present a wide range of results that compare the polarizabilities from
\ISApol with those from \cDF and \SRLO, and demonstrate that the new
models are superior in many ways. Finally we compare the dispersion energies
from {\em localized} \ISApol models with those from SAPT(DFT).
We end with an outlook on the scope and power of this method.

\section{Theory}
\label{sec:theory}
The frequency-dependent polarizability tensors can be defined from the 
frequency-dependent density susceptibility (FDDS) function and the 
multipole moment operators (or any one-electron operators\cite{MisquittaS06})
as follows
\begin{align}
  \pol{}{tu} &= \iint \oQ{}{t}(\rr) \fdds \oQ{}{u}(\rp) \dr \drp.
           \label{eq:pol}
\end{align}
where $\oQ{}{t}$ is the (real) multipole moment operator of index $t$
where the index (rank and component) is expressed in the compact
notation of Stone \cite{Stone:book:13}: $t=00,10,11c,11s,\cdots$.
The FDDS describes the linear response of the electron density to a
frequency-dependent perturbation and can be written in sum-over-states form
as 
\begin{equation}
  \fdds = \sum_{n \ne 0} 
             \frac{2\omega_{n}}{\hslash (\omega_{n}^{2} - \omega^{2})}
             \langle 0|\hat{\rho}(\rr)|n \rangle
             \langle n|\hat{\rho}(\rp)|0 \rangle,
             \label{eq:FDDSdef}
\end{equation}
where $\hat{\rho}(\rr) = \sum_{k} \delta(\rr - \rr_{k})$ is the 
electron density operator and $k$ runs over the electrons in the system.

To achieve a partitioning of the total molecular polarizability, \eqrff{eq:pol},
into contributions from the AIM domains we define a unit function:
\begin{align}
  \mathbf{I}(\rr) &= \sum_a \left( \frac{\w{a}(\rr)}{\sum_c \w{c}(\rr)} \right)
                     \nonumber \\
                  &= \sum_a \p{a}(\rr),
  \label{eq:unit-func}
\end{align}
where $\p{a}(\rr)$ is the probability of a quantity being associated 
with AIM $a$ at point $\rr$.
With two such unit functions we can define the distributed form of the FDDS
as follows:
\begin{align}
  \fdds &= \mathbf{I}(\rr) \ \fdds \ \mathbf{I}(\rp) \nonumber \\
        &= \sum_a \sum_b \left( \p{a}(\rr) \ \fdds \ \p{b}(\rp) \right) \nonumber \\
        &= \sum_a \sum_b \fddsdist{ab}.
  \label{eq:ISApol-fdds}
\end{align}
Notice that the FDDS, being a two point function, is partitioned into 
contributions from pairs of sites.
Having thus partitioned the FDDS, we can now define the distributed, non-local
polarizabilities as
\begin{align}
  \pol{ab}{tu} &= \iint \oQ{}{t}(\rr-\RR_a)\ \fddsdist{ab}\ \oQ{}{u}(\rp-\RR_b) \dr \drp,
   \label{eq:ISApol-pol}
\end{align}
where the multipole moment operators are now defined using the centres of
sites $a$ and $b$. These are the distributed multipole operators, for
which we will 
also use the notation $\oQ{a}{t}(\rr) \equiv \oQ{}{t}(\rr-\RR_a)$.

\subsection{A simplified and flexible BS-ISA algorithm}
\label{sec:new-bs-isa}

In the BS-ISA algorithm \cite{MisquittaSF14} we represent
the ISA atomic density for site $a$, \R{a}, in terms of an appropriate local,
atomic basis set:
\begin{align}
  \R{a}(\rr) = \sum_k c_k^a \ \xi_k^a(\rr),
  \label{eq:rhoA-expansion}
\end{align}
where the $\xi_k^a$ are basis functions associated with site $a$
and the coefficients $c_k^a$ are determined by minimizing an 
appropriate ISA functional (see below).
The piece-wise continuous shape function $\wT{a}$ is defined as
\begin{align}
  \wT{a}(\rr) = 
    \begin{cases}
      \w{a}(\rr)  & \text{if} \ |\rr| \leq r^a_0 \\
      \wL{a}(\rr) & \text{otherwise}.
    \end{cases}
    \label{eq:tail-fix}
\end{align}
where the transition radius $r^a_0$ is defined appropriately \cite{MisquittaSF14}.
The short-range form $\w{a}$ is given by a basis expansion:
\begin{align}
  \w{a}(\rr) &= \sum_{k \in \text{s-func}} c_k^a \; \xi_{k,{\rm s}}^{a}(\rr),
  \label{eq:w-expansion}
\end{align}
where the basis set consists of s-type functions taken from the basis used for the 
atomic expansion given in \eqrff{eq:rhoA-expansion}.
The long-range form of the shape function is given by 
\begin{align}
  \wL{a}(\rr) = A_{a} \exp{(-\alpha_{a} |\rr - \RR_{a}|)}, 
    \label{eq:wL}
\end{align}
where the constants $A_{a}$ and $\alpha_{a}$ are obtained self-consistently
\cite{MisquittaSF14}. 
As we have previously explained, the purpose of this piece-wise 
definition of the shape function is to enforce the exponential 
decay of the ISA atomic densities, which is difficult to obtain with
Gaussian basis sets as the very diffuse basis functions needed to model 
the long-range density tails tend to lead to numerical instabilities.
Using $\wL{a}$ allows us to obtain an exponential decay without
needing to use very diffuse basis functions.

The ISA solutions are then be obtained from an iterative process, where,
at each step of the iterations a suitable functional is minimized. 
One of these is the \ISAfunc{A} functional which is the default
in the \CamCASP{} program. 
A computationally important feature of the \ISAfunc{A} functional 
is that it can be minimized with \order{N} computational cost, where
$N$ is the number of ISA {\em sites} in the system. This is possible 
as the \ISAfunc{A} functional can be written as the sum of sub-functionals:
\begin{align}
  \ISAfunc{A} &= \sum_{a} \left\| 
         \biggl( \R{a} - \rho \frac{\wT{a}}{\sum_b \wT{b}} \biggr)^2
                                 \right\|, \nonumber \\
              &= \sum_{a} \ISAfunc{A}^{a},
  \label{eq:isa-A}
\end{align}
where each of the sub-functionals, $\ISAfunc{A}^{a}$ can be minimized
independently of the others. 
Importantly, the total density $\rho$ used in this functional is obtained 
via density fitting \cite{MisquittaSF14}; this is needed to 
reduce the computational scaling to \order{N}, and it also simplifies the
integrals needed.

However in the original implementation, minimizing the \ISAfunc{A} functional 
tended to lead to unacceptable inaccuracies in the ISA AIM densities; in particular
the total charge of the system was often not conserved, with differences
of $0.01$e often encountered. Also, higher ranking molecular multipoles would 
not be well reproduced.
Consequently we combined the \ISAfunc{A} functional with the density-fitting
functional to result in a hybrid DF-ISA algorithm. 
This algorithm involved a single parameter that controlled
the relative weights given to each scheme, with a $90$\% weighting of the
DF functional being recommended. 
While the results were better, there were two problems: 
(1) the new method had a computational scaling of $\order{N^3}$,
and, 
(2) despite the mixture of the density-fitting and ISA functionals, there
was still an overall loss in accuracy which resulted in small residual errors
in the electrostatic energies computed from the DF-ISA algorithm
compared with reference energies from SAPT(DFT).

The primary reason for the inaccuracy of the original algorithm was that the ISA atomic
basis sets were constructed from the auxiliary basis used in the density fitting, 
and this inextricably linked the two basis sets. 
This placed limits on both basis sets, and therefore resulted in inaccuracies both
in the fitted density and in the ISA solutions.
This restriction in the basis sets was required for technical reasons associated
with the implementation of the BS-ISA algorithm in version 5.9 of the \CamCASP{} program.
It was because of these inaccuracies that we needed to use the 
more computationally demanding DF-ISA algorithm.

In the present algorithm implemented in \CamCASP 6.0 we have removed these 
restrictions by introducing a third, independent, atomic basis set in the 
\CamCASP program which now contains the following bases:
\begin{itemize}
  \item {\em The main basis}: used for the molecular orbitals.
  \item {\em The auxiliary basis}: used for the density fitting. This 
    basis may use either Cartesian or spherical GTOs.
  \item {\em The atomic basis sets}: used for the ISA atomic expansions.
    This basis set {\em must} use spherical GTOs, but is otherwise 
    independent from the above basis sets. The atomic basis sets 
    can therefore be increased in size if needed and placed on 
    arbitrary sites, or removed from some sites. 
\end{itemize}
With this change, we are now able to control the variational flexibility of
the ISA solution independently of that of the density fitting. 
As the ISA expansions are known to require an increased variational flexibility 
compared with the density fitting, we can now use larger basis
for the ISA expansions, thereby leading to overall higher accuracies
with functional \ISAfunc{A}; there is no longer a need to use the 
DF-ISA algorithm. 
This not only restores the \order{N} computational scaling of the algorithm,
but also allows us to use Cartesian GTOs in the density-fitting step,
thereby significantly reducing the errors in the fitted density.

In addition, we have made improvements to the way in which distributed 
molecular properties are extracted using the ISA solutions.
Previously, distributed molecular properties such as the multipole moments 
were defined in terms of the ISA atomic expansions $\R{a}$:
\begin{align}
  \Q{a}{t} &= \int \oQ{a}{t}(\rr) \R{a}(\rr) \dr,
    \label{eq:distmom-ISA}
\end{align}
where $\Q{a}{t}$ is the (real) distributed multipole moment of index $t$
for site $a$.
In the new scheme we instead use the expression
\begin{align}
  \Q{a}{t} &= \int \oQ{a}{t}(\rr) \rho(\rr) \frac{\wT{a}(\rr)}{\sum_b \wT{b}(\rr)} \dr 
               \nonumber \\
           &= \int \oQ{a}{t}(\rr) \rho(\rr) \p{a}(\rr) \dr.
    \label{eq:distmom-ISA-GRID}
\end{align}
This expression is formally identical to \eqrff{eq:distmom-ISA}, but as 
\eqrff{eq:stockholder-partitioning} is never an identity, 
the latter expression is usually more accurate. 
We refer to multipole moments computed with \eqrff{eq:distmom-ISA-GRID}
as the ISA-GRID moments.

\section{Numerical implementation}
\label{sec:numerical}

For single-reference wavefunctions, such as those from Hartree--Fock (HF)
and Kohn--Sham density functional theory (DFT), the FDDS can be evaluated
using coupled linear-response theory and is expressed as a sum over
occupied and virtual single-particle orbitals and eigenvalues as
\begin{equation}
   \fdds = \sum_{iv,i'v'} C_{iv,i'v'}(\omega)
               \phi_{i}(\rr) \phi_{v}(\rr) \phi_{i'}(\rp) \phi_{v'}(\rp),
               \label{eq:fdds}
\end{equation}
where the subscripts $i$ and $i'$ ($v$ and $v'$) denote occupied (virtual) 
molecular orbitals, $\phi$ are the single-particle orbitals, and the 
frequency-dependent coefficients $C_{iv,i'v'}(\omega)$ 
are defined in terms of the electric and magnetic Hessians 
\cite{Casida95,ColwellHL96,MisquittaS06}.
Using density fitting \cite{DunlapCS79,Dunlap00a,JamorskiCS95} 
we express the transition densities in terms of an auxiliary basis $\chi_k$:
\begin{align}
  \phi_{i}(\rr) \phi_{v}(\rr) &= \sum_k D_{iv,k}\, \chi_k(\rr),
  \label{eq:df}
\end{align}
and this allows us to write the FDDS as \cite{MisquittaJS03,HesselmannJS05,MisquittaS06}
\begin{align}
  \fdds \approx \sum_{k,l} \tilde{C}_{kl}(\omega) \ \chi_k(\rr) \ \chi_l(\rp),
  \label{eq:df-fdds}
\end{align}
where the $\tilde{C}_{kl}(\omega)$ are the transformed coefficients which
are defined as
\begin{align}
  \tilde{C}_{kl}(\omega) &= \sum_{iv,i'v'} D_{iv,k} C_{iv,i'v'}(\omega) D_{i'v',l}.
                \label{eq:Ctilde}
\end{align}

Using the density-fitted form of the FDDS in \eqrff{eq:ISApol-pol} we get
\begin{align}
  \pol{ab}{tu} &= \sum_{k,l} \tilde{C}_{k,l}(\omega) 
            \iint \oQ{a}{t}(\rr)\ \p{a}(\rr)\ \chi_{k}(\rr) \chi_{l}(\rp)
             \p{b}(\rp)\ \oQ{b}{u}(\rp) \dr \drp \nonumber \\
         &= \sum_{k,l} \tilde{C}_{k,l}(\omega) 
               \left( \int \oQ{a}{t}(\rr)\ \p{a}(\rr)\ \chi_{k}(\rr) \dr \right) \nonumber \\
         &  ~~~~~~~~~~~~~~~~~~   \times
               \left( \int \oQ{b}{u}(\rp)\ \p{b}(\rp)\ \chi_{l}(\rp) \drp \right)
             \nonumber \\
         &= \sum_{k,l} \Q{a}{t,k}\ \tilde{C}_{k,l}(\omega)\ \Q{b}{u,l},
   \label{eq:ISApol-dist-pol}
\end{align}
where in the last step we have defined the distributed multipole moment integrals for
sites $a$/$b$ and auxiliary basis functions $k$/$l$:
\begin{align}
  \Q{a}{t,k} &= \int \oQ{a}{t}(\rr)\ \p{a}(\rr)\ \chi_{k}(\rr) \dr \nonumber \\
             &= \int \oQ{a}{t}(\rr)\ \frac{\wT{a}(\rr)}{\sum_b \wT{b}(\rr)}
                     \ \chi_{k}(\rr) \dr.
   \label{eq:Q-int-aux}
\end{align}
Notice that these multipole integrals are analogous with those used to define
the ISA-GRID multipole moments shown in \eqrff{eq:distmom-ISA-GRID}.
This is the \ISApol model for distributed frequency-dependent non-local 
polarizabilities.

In the \cDF and \SRLO methods the distribution is achieved via the auxiliary
basis functions themselves \cite{MisquittaS06,RobS13a}. These methods are linked
to the \ISApol algorithm by setting the probability functions $\p{a}(\rr) = 1$
and limiting the sum over $k$/$l$ in \eqrff{eq:ISApol-dist-pol} to include only 
those auxiliary functions on sites $a$/$b$. 
This has the advantage of simplicity, but disadvantage that the results are
dependent on the auxiliary basis set \cite{MisquittaS06}.
In the \ISApol approach the distributed polarizabilities are uniquely 
defined for a given set of probability functions $\p{a}$, and as
we know that the ISA solutions are unique \cite{LillestolenW09,BultinckCN09},
we should expect that the \ISApol algorithm leads to unique distributed
polarizabilities. We shall demonstrate this below.

\subsubsection{Linearising the algorithm: Issues}
\label{sec:linear-scaling}


Once the frequency-dependent coefficients $\tilde{C}_{kl}(\omega)$ have been
calculated, the evaluation of $\pol{ab}{tu}$ using \eqrff{eq:ISApol-dist-pol}
for a given pair of sites $a,b$ and angular momenta $t,u$
scales as $\order{M^2}$ where there are $M$ auxiliary basis functions
in the system. 
If $l$ is the maximum angular momentum for which distributed polarizabilities
to be computed and $N$ is the number of sites in the system, then
there are $\order{N^2\, l^4}$ non-local polarizabilities, so the total scaling
of the calculation is $\order{l^4\, N^2\, M^2}$.
If we assume on the average $m$ auxiliary basis functions per site, then 
$M = mN$, so the computational scaling is $\order{l^4\, m^2\, N^4}$, that is, it
scales as the fourth power as the number of sites. 
While the scaling is not necessarily unfavourable, the pre-factor, $l^4\, m^2$,
can easily be of the order $10^6$, thereby making this calculation 
computationally burdensome, though it can be trivially parallelized over the
pairs of sites $a,b$.

The distributed multipole integral in the auxiliary basis defined
in \eqrff{eq:Q-int-aux} must be evaluated numerically, on a grid due to 
the ISA probability function $\p{a}$. This function is defined as the 
ratio of the ISA shape functions which makes analytic evaluation unfeasible,
but these are themselves piece-wise continuous, so numerical evaluation 
is mandatory.
As the numerical integration grid size scales with the number of atoms in the
system, the evaluation of the $\Q{a}{t,k}$ integrals using \eqrff{eq:Q-int-aux}
would incur a computational cost scaling as $\order{l^2\, m\, n_g\, N^3 }$,
where $n_g$ is the average number of grid points per atom, that is,
the scaling is $\order{N^3}$ with number of atoms.
As we need fairly dense grids, particularly in the angular coordinates, to 
converge the higher ranking multipole moment integrals, the pre-factor 
$l^2\, m\, n_g$ can be as large as $10^7$.
This can make the evaluation of these integrals a significant computational 
cost, and even though this evaluation needs to be done only once in a calculation,
it would be advantageous if the scaling could be reduced.

Fortunately both of these computational costs can be reduced using locality
enforced by defining neighbourhoods for each site in the system \cite{MisquittaSF14}. 
We define the neighbourhood \neighbourhood{a} of site $a$ as site $a$ itself and 
all other sites whose auxiliary basis functions overlap with those of site $a$ 
within a specified threshold. 
Now consider how the neighbourhood \neighbourhood{a} can be used to reduce the
computational cost of the multipole moment integrals $\Q{a}{t,k}$ for site $a$:
\begin{itemize}
  \item {\em Integration grids}: Rather than spanning all atoms in the system,
    the grids are based on sites in \neighbourhood{a}.
  \item {\em Probability function evaluation}: $\p{a}$ includes a sum over all
    sites in the system, but this sum can be restricted to go over only 
    sites in \neighbourhood{a}.
  \item {\em Auxiliary basis function $k$}: $\Q{a}{t,k}$ is evaluated only for 
    those $k$ that belong to sites in \neighbourhood{a} and is set to zero otherwise.
\end{itemize}
With these three changes, the computational cost of evaluating the multipole
integrals is reduced to \order{N}.

In a similar manner the cost of evaluating \eqrff{eq:ISApol-dist-pol} 
is reduced to $\order{N^2}$ by restricting the sum over auxiliary basis function
indices $k$ and $l$ to include only those functions from sites in the 
neighbourhood of sites $a$ and $b$, respectively: 
\begin{align}
  \pol{ab}{tu} &= \sum_{a' \in \neighbourhood{a}} \sum_{k \in a'}
                  \sum_{b' \in \neighbourhood{b}} \sum_{l \in b'}
          \Q{a}{t,k}\ \tilde{C}_{k,l}(\omega)\ \Q{b}{u,l}.
   \label{eq:ISApol-dist-pol-local}
\end{align}

At present we use the same neighbourhood definition for the integration grids,
ISA probability functions, and auxiliary basis functions. This may not be ideal
as it is quite possible that efficiency gains may be obtained by using different
definitions for the three. We have yet to explore such a possibility.

There are limitations to the use of neighbourhoods to achieve linearity in 
computational scaling: for heavily delocalized systems such as the
$\pi$-conjugated molecules the neighbourhoods may need to be increased
in order to achieve sufficient accuracy in the polarizabilities.
In this case, using neighbourhoods that are too small leads
to increased charge-conservation errors in the BS-ISA solution,
and to sum-rule violations in the charge-flow \cite{Stone:book:13}
contributions to the non-local polarizabilities.

%
%

\subsection{Localization of the non-local polarizabilities}
\label{sec:localization}

The main focus of this paper is not the non-local polarizabilities
defined in \eqrff{eq:ISApol-dist-pol}, but rather the 
{\em localized} distributed polarizability models that can be 
derived from these using techniques described in detail in some of
our previous publications \cite{MisquittaS08a,MisquittaSP08}.
This is not to diminish the importance of the non-local polarizability
models, indeed these models are essential for heavily delocalized 
systems, and in low dimensional systems leads to van der Waals interactions that
cannot be replicated by any local model \cite{MisquittaSSA10,LiuAD11,MisquittaMDSN14}.
However it is the local models that are commonly used, so for very pragmatic
reasons we will focus on these here.

Local polarizability models are an approximation, but one that often turns
out to be reasonable, particularly for insulators for which electron correlations
are largely local.
In the \WSM algorithm \cite{MisquittaS08a,MisquittaSP08} we have defined a
means for converting any non-local polarizablity model into an effective
local one using two transformation steps:
\begin{itemize}
  \item {\em Multipolar localization}: In the two-step localization scheme
     that forms part of the \WSM model we first transform away the 
     non-local contributions using a multipole expansion 
     \cite[\S9.3.3]{Stone:book:13}.
     We have explored two schemes for this purpose: the method of 
     LeSueur \& Stone \cite{LeSueurS94} and that of 
     Wheatley \& Lillestolen \cite{WheatleyL08}. Of these, the latter has the
     advantage that the non-local terms are localized along the molecular bonds
     and should result in better convergence of the resulting model.
     However either of these localization procedures lead to a degradation
     in the convergence of the resulting polarizability expansion.
  \item {\em Constrained refinement}: In this step the multipolar localized
     polarizability models are {\em refined} to reproduce the point-to-point
     polarizabilities \cite{WilliamsS03}${}^,$\cite[\S9.3.2]{Stone:book:13} 
     computed on a pseudo-random set of points surrounding the molecule.
     The idea here is to use the local polarizabilities from
     the first step as prior values, and allow them to relax using constraints to
     keep them close to their original values. 
\end{itemize}
These steps can be performed for polarizabilities at any frequency.
One of the features of this approach is that at the refinement stage
symmetries can be imposed, and if needed, models may be simplified. 
The \WSM procedure ensures that the best resulting model is obtained. 

In the original \WSM model we relied on non-local polarizabilities from the
\cDF algorithm as the starting point. This did not always work out well as
the multipolar localized models often contained terms with unphysical values
which would change by a considerable amount in the refinement stage. 
For this reason the constraints we recommended \cite{MisquittaSP08} were 
weak for the dipole-dipole polarizabilities, and completely absent for the 
higher ranking terms. The lack of constraints for the higher ranking terms
was simply a recognition that our prior values were simply too unreliable. 
Looked at another way, the final polarizability models depended quite strongly
on the kinds of constraints used.

Here we use the \ISApol non-local polarizabilities as input to the \WSM algorithm.
From empirical observation we know that the multipolar localized models 
are already good and only relatively small changes occur on refinement.
However the refinement step does still improve the localized models, so we
continue to use it, but this time with much stricter constraints.
Referring to eq.(36) in Ref.~\citenum{MisquittaS08a} 
(see also eq. {9.3.13} in Ref.~\citenum{Stone:book:13}),
we now define the constraint matrix to be
\begin{align}
  g_{kk'} &= \delta_{kk'} \frac{w_0}{1+(p^0_k)^2},
    \label{eq:refinement-weights}
\end{align}
where $k$/$k'$ is a model parameter index (these label the polarizabilities),
$\delta_{kk'}$ is the Kroneker-delta function, $w_0$ is a constant, and
$p^0_k$ is the reference value of the parameter (that is, the local polarizability)
obtained from the multipolar step.
We use $w_0 = 10^{-3}$ for calculations on the larger systems, but for smaller
systems, where there is sufficient data in the point-to-point polarizabilities
to yield a meaningful refinement of even the higher-ranking polarizabilities,
the constraints may be relaxed using $w_0 = 10^{-5}$.

It may seem paradoxical to use constraints of any kind if the refinement step
does not alter the multipolar localized \ISApol model by much. 
The reason for the use of constraints is that in a mathematical optimization it is
possible for parameters to alter without a meaningful change in the cost-function.
The constraints prevent this kind of mathematical wandering of parameters,
particularly for large systems for which we rarely have enough data in the
point-to-point polarizabilities to act as natural constraints to the parameters.

\section{Numerical details}

All SAPT(DFT) calculations have been performed using
the \CamCASP{5.9} program \cite{CamCASP} with orbitals and energies computed using the
\DALTON program \cite{DALTON2} with a patch installed from the \SAPT code.
The Kohn--Sham orbitals and orbital energies were computed using an
asymptotically corrected PBE0 \cite{AdamoB99a} functional with 
Fermi--Amaldi (FA) long-range exchange potential \cite{FermiA34} and
the Tozer \& Handy splicing scheme.
Linear-response calculations and \ISApol polarizabilities were performed using the
same functional but with a developer's version of \CamCASP{6.0}.
The kernel used in the linear-response calculations is the hybrid ALDA+CHF kernel
\cite{MisquittaPJS05b,MisquittaS06}
which contains 25\% CHF (coupled Hartree--Fock) and 75\% ALDA
(adiabatic local-density approximation). This kernel is constructed within
the \CamCASP{} code. The PW91 correlation functional \cite{PerdewW92} 
is used in the ALDA kernel.

The shift needed in the asymptotic correction has been computed self-consistently
using the following ionization potentials:
thiophene: 0.326 a.u.\ \cite{Lias00}; 
pyridine: 0.3488 a.u.\ \cite{MisquittaS16};
water: 0.4638 a.u.\ \cite{Lias00};
methane: 0.4634 a.u.\ \cite{Lias00}.
The vibrationally averaged molecular geometry was used for water \cite{MasS96}
and methane \cite{NakataK86,Akin-OjoS05} molecules, the pyridine geometry has been taken 
from Ref.~\citenum{MisquittaS16}, and the thiophene geometry has been 
obtained by geometry optimization using the PBE0 functional and
the cc-pVTZ basis \cite{KendallDH92} with the \NWChem program \cite{Valiev:NWChem:10}. 

The SAPT(DFT) calculations use two kinds of basis sets: the main basis, used in the
density-functional calculations, is in the \MCpBS basis format,
that is, with mid-bond and far-bond functions, and the 
auxiliary basis used for the density fitting is in the \DCpBS format.
The following main/auxiliary basis sets were used for the systems studies in this
paper:
\begin{itemize}
  \item {\em Methane dimer, water dimer, methane..water complex}: 
     main basis: aug-cc-pVTZ with 3s2p1d mid-bond set,
     and auxiliary basis: aug-cc-pVTZ-RI basis with 3s2p1d-RI basis.
  \item {\em Pyridine dimer}: main basis: Sadlej-pVTZ \cite{Sadlej88} 
     with a 3s2p1d mid-bond set \cite{BukowskiSJJSKWR99},
     and auxiliary basis: aug-cc-pVTZ-RI basis \cite{WeigendKH02} 
     with 3s2p1d-RI basis. 
\end{itemize}

The \ISApol calculation is preceded by a BS-ISA calculation which is subsequently
fed into the distributed polarizability module in \CamCASP{}. As described
in Ref.~\citenum{MisquittaSF14}, the ISA expansions use basis sets created
from a special set of s-type functions with higher angular momentum functions
taken from a standard resolution of the identity (RI) fitting basis. 
We have used the following combinations of basis sets for the calculations 
reported in this paper:
\begin{itemize}
  \item {\em The methane and water molecules}:
    main basis: d-aug-cc-pVTZ (spherical); 
    auxiliary basis: aug-cc-pVQZ-RI (Cartesian) with ISA-set2 with s-functions on the 
      hydrogen atoms limited to a smallest exponent of $0.25$ a.u.\;
    atomic basis: like the auxiliary basis, but with spherical GTOs.
  \item {\em The pyridine molecule}:
    main basis: d-aug-cc-pVTZ (spherical); 
    auxiliary basis: aug-cc-pVQZ-RI (Cartesian);
    atomic basis: aug-cc-pVQZ-RI (spherical) with ISA-set2.
\end{itemize}
For these three molecules we used the \ISAfunc{A} functional for the ISA calculations,
but for the thiophene molecule we used the older `A+DF' algorithm in which we first 
converge the ISA solution using the \ISAfunc{A} functional, and subsequently 
use the DF+ISA algorithm with $\zeta=0.1$, that is, with a weighting of 10\% given to 
\ISAfunc{A} and 90\% to the density-fitting functional.
As we have discussed in \secrff{sec:new-bs-isa},
the DF+ISA algorithm places restrictions on the auxiliary basis set, so the 
basis sets used for the thiophene molecule are different, with the auxiliary 
and atomic basis sets being the same.
For thiophene we have reported results using three kinds of main basis sets:
for the aug-cc-pVDZ and aug-cc-pVTZ main basis sets, we have used
an auxiliary basis consisting of ISA-set2 s-type functions with higher
angular functions taken from the aug-cc-pVTZ-RI basis with spherical GTOs,
and for the aug-cc-pVQZ main basis we have used an auxiliary basis
consisting of s-functions from the ISA-set2 basis with higher angular terms
from the aug-cc-pVQZ-RI basis also using spherical GTOs.
We have not used the aug-cc-pVDZ-RI basis as it is not large enough for
an ISA calculation.

\section{Results}
\label{sec:results}

Although the non-local polarizability models are fundamental, these
are also, at present, of high complexity and are not suitable for
most applications. So while we assess some features of the \ISApol
non-local polarizability models, we will here be primarily concerned
with the localized models.

\subsection{Convergence with rank}
\label{sec:conv-rank}

The assessment of the polarizability models is complicated by the fact that
there is no pure polarization energy defined in SAPT or SAPT(DFT):
the second-order induction energy in these methods contains both a polarization and
a charge-transfer contribution. While it is possible to separate these, for example
using regularized SAPT(DFT) \cite{Misquitta13a}, we inevitably then encounter 
the problem of damping \cite{MisquittaS08a,MisquittaS16}.
An elegant solution to the first problem is to compute the polarization energy
of the molecule interacting with a point charge probe. 
This has the advantage that the energies can be easily displayed on a surface 
around the molecule, and as reference energies can be easily computed
using the \CamCASP{} program, it is relatively straightforward to make 
comparisons of the model and reference energies and visualise the 
differences on the molecular surface.

There is however still the issue of the damping, and we have chosen to use a
simple proposal: a single parameter Tang--Toennies \cite{TangT84} 
damping model is used, and the damping parameter is determined by requiring that
the mean signed error (MSE) of the damped model energies against the reference 
SAPT(DFT) energies is as small as possible.
We have studied three series of polarization models for each of the 
\ISApol and \cDF distribution algorithms: the non-local, and localized
isotropic and anisotropic models. 
We have determined a polarization damping parameter for each of the six series
of models from the highest ranking model in the series; this parameter
is then fixed for all lower ranking models in the series.
For the \ISApol models the damping parameters are $1.57$, $1.50$ and $1.51$ a.u.\
for the non-local, local (anisotropic), and local (isotropic) models, respectively,
while the corresponding damping parameters for the \cDF models are
$1.32$, $1.49$ and $1.61$ a.u.

%

\begin{figure*}
  \includegraphics[width=0.9\textwidth]{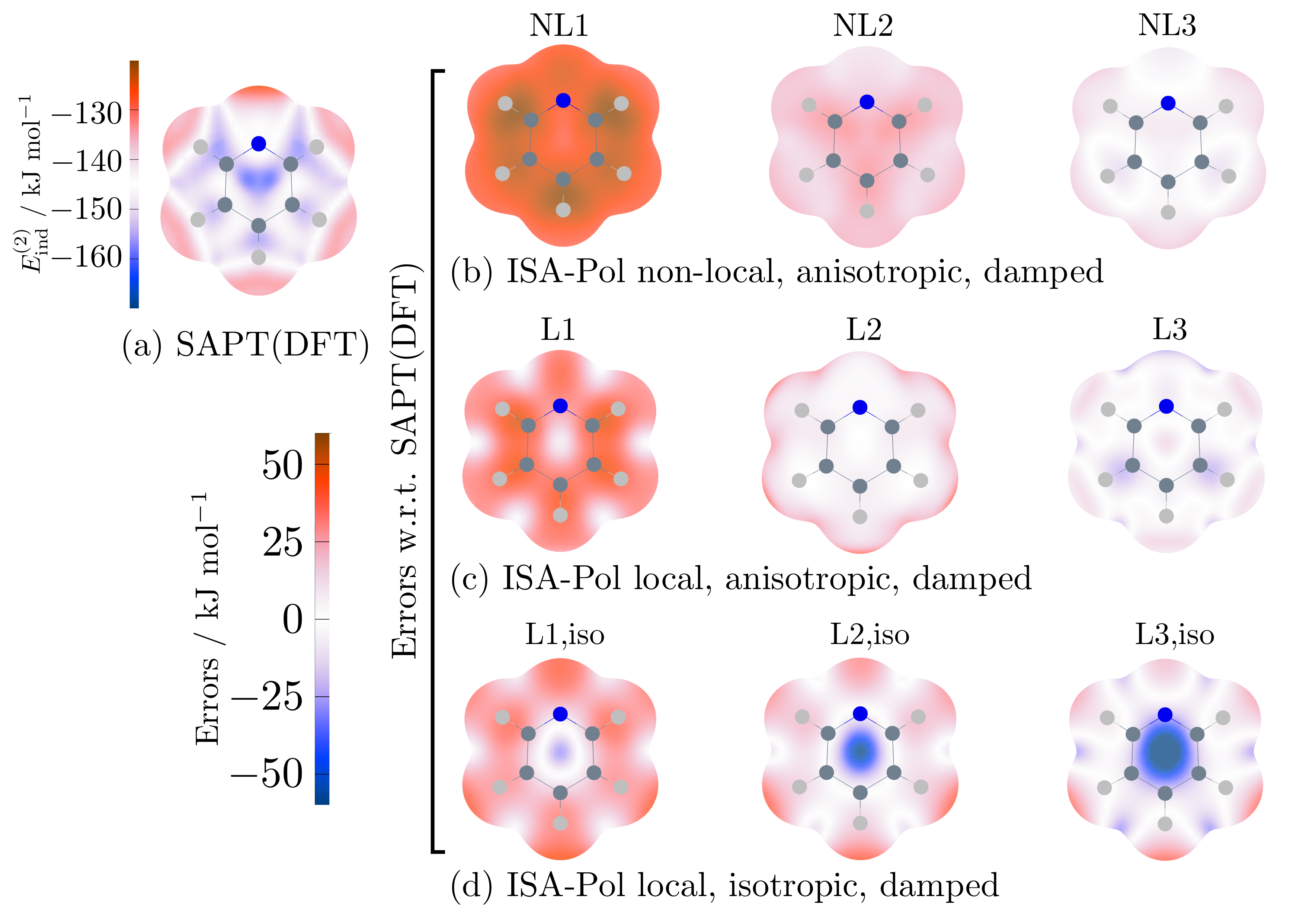}
  \caption[Comparison of damped \ISApol polarization models]{
      Comparison of the polarization energies for pyridine interacting with a 
      $+1$e point charge on the $0.001$e isodensity surface of pyridine.
      In panel (a) we visualise the reference SAPT(DFT) second-order induction 
      energies. In the other panels we visualise the errors made by various 
      {\em damped} polarization models from the \ISApol algorithm: 
      (b) non-local models, (c) localized, anisotropic models, and 
      (d) localized, isotropic models. The maximum rank of the polarizability
      model is indicated.
  \label{fig:ISApol-pol-models-damped}
  }
\end{figure*}

\begin{figure*}
  \includegraphics[width=0.9\textwidth]{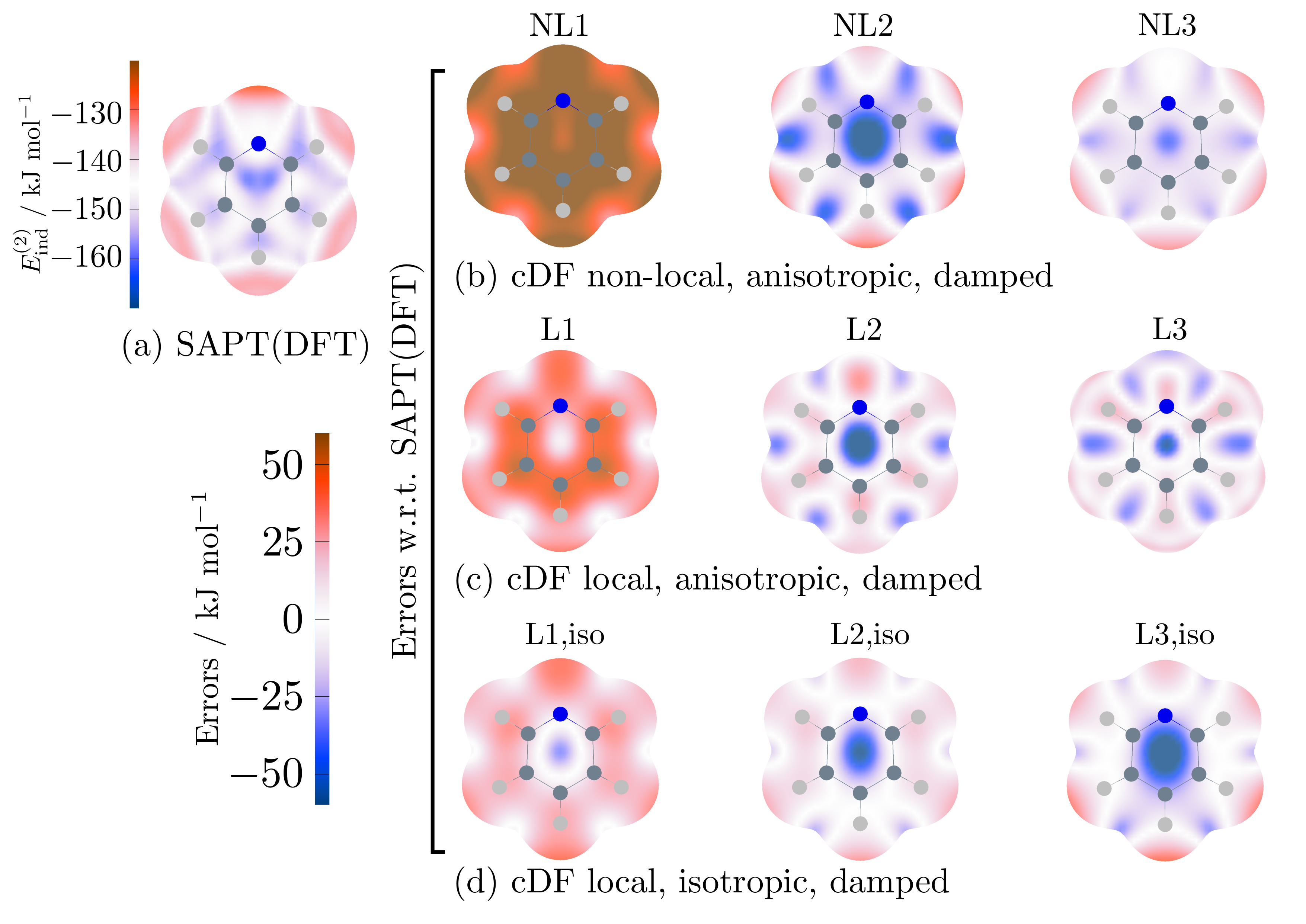}
  \caption[Comparison of damped \cDF polarization models]{
      Comparison of the polarization energies for pyridine interacting with a 
      $+1$e point charge on the $0.001$e isodensity surface of pyridine.
      In panel (a) we visualise the reference SAPT(DFT) second-order induction 
      energies. In the other panels we visualise the errors made by various 
      {\em damped} polarization models from the \cDF algorithm: 
      (b) non-local models, (c) localized, anisotropic models, and 
      (d) localized, isotropic models. The maximum rank of the polarizability
      model is indicated.
  \label{fig:cDF-pol-models-damped}
  }
\end{figure*}


In \figrff{fig:ISApol-pol-models-damped} we have displayed the reference 
SAPT(DFT) polarization energies for the pyridine molecule interacting with
a $+1$e point charge probe. The energies are displayed on a $10^{-3}$ isodensity
surface computed using the \CamCASP{} program. 
The resulting polarization energies are uncharacteristically large, due both to this
choice of surface (which corresponds approximately to the van der
Waals surface) and to the large size of the charge: typical local
charges in atomic systems will usually be half as much.
Also shown in \figrff{fig:ISApol-pol-models-damped} are the errors made by the 
damped polarization models against the reference energies. 
Consider first the non-local models: the positive errors made by the NL1 
model indicate an underestimation of the polarization energy.
The agreement with the reference energies gets progressively and systematically
better as the maximum rank increases through 2 to 3. Results for the
NL4 model (the maximum rank of the non-local models, and also the most
accurate for the choice of damping) are not shown.
The localized, anisotropic models exhibit similar errors, 
but the localized, isotropic models show larger variations in the errors made. 
In particular, these models shown an underestimation of the polarization near
the hydrogen and nitrogen atoms, and a large overestimation of the polarization
in the centre of the ring. 
This is due to the simplicity of the isotropic models: the
polarizability of an anisotropic system like pyridine cannot be correctly 
modelled everywhere using isotropic AIM polarizabilities. 
As with the distributed multipole moments \cite{MisquittaSF14}, the 
ISA AIMs lead to polarization models with better convergence with
increasing rank and fewer artifacts
in both the non-local and local models.

In \figrff{fig:cDF-pol-models-damped} are shown similar results, this time for the
models from the \cDF algorithm. These differ from the \ISApol models in 
important ways: first of all the errors are larger, even for the 
non-local models, but perhaps more importantly, the {\em variations} in
the errors are much larger for all models. 
It is the latter that is the bigger concern for model building, as 
variations in errors arise from to position and angle dependent variations
in the quality of the model, leading to unreliable predictions.

\section{Convergence with basis of the localized models}
\label{sec:conv-basis}

The next question we need to address is the basis set convergence
of the \ISApol models. We will not discuss the performance of the non-local
or local anisotropic models here as it is difficult to display the 
data contained in these models in a meaningful and concise manner.
Instead we will focus on the local, isotropic models.

The construction of a local, isotropic (frequency-dependent) polarizability 
model begins with the multipolar localization (see \secrff{sec:localization})
of the \ISApol non-local model. 
This results in an {\em anisotropic}, local model which has not yet been 
refined against the point-to-point polarizabilities.
The isotropic model may now be obtained in one of three ways:
\begin{itemize}
  \item Directly from the unrefined anisotropic model by retaining only the
    isotropic part of the polarizabilities.
  \item By refining this isotropic model using the point-to-point polarizabilities.
  \item By refining the anisotropic model as described in \secrff{sec:localization}
    and {\em subsequently} retaining only the isotropic part of the polarizabilities.
\end{itemize}
The second and third options should, in principle, lead to more accurate models.
These two approaches lead to similar, but not identical local, isotropic
polarizability models. By refining the isotropic models (the second option)
we ensure that the resulting isotropic models are the most accurate possible
given the limitations imposed. But while this approach may be applicable to 
small systems for which the isotropic approximation may be valid, it will 
fail for strongly anisotropic systems for which the third approach may be 
more appropriate. 
We have used the second method to obtain the isotropic polarizability models
discussed in this paper.

\begin{table}[ht]
   \newcolumntype{d}{D{.}{.}{2}}
   \begin{tabular}{llddd}
     \toprule
       Site  & $l$ &\multicolumn{1}{r}{\text{aDZ}}
                               &\multicolumn{1}{r}{\text{aTZ}}
                                         &\multicolumn{1}{r}{\text{aQZ}}\\
     \midrule
       C1    &  1  &     7.39  &    7.40  &    7.36  \\
             &  2  &    23.48  &   27.40  &   28.21  \\
             &  3  &   458.12  &  553.61  &  579.94  \\
     \midrule
       C2    &  1  &    10.53  &   10.62  &   10.64  \\
             &  2  &    32.58  &   37.75  &   38.85  \\
             &  3  &   654.79  &  766.85  &  806.50  \\
     \midrule
       S     &  1  &    16.74  &   17.05  &   17.17  \\
             &  2  &    97.09  &  109.74  &  115.85  \\
             &  3  &  1449.90  & 1859.26  & 2232.79  \\
     \midrule
       H1    &  1  &     2.18  &    2.17  &    2.18  \\
             &  2  &     4.41  &    4.44  &    4.74  \\
             &  3  &   -17.88  &   -6.67  &   -0.36  \\
     \midrule
       H2    &  1  &     1.53  &    1.47  &    1.46  \\
             &  2  &     4.00  &    4.07  &    4.16  \\
             &  3  &   -11.64  &   -2.93  &    5.42  \\
     \bottomrule
   \end{tabular}
   \caption{
     Localized, isotropic polarizabilities for the symmetry-distinct
     sites in the thiophene molecule computed with the WSM algorithm 
     starting from \ISApol non-local polarizabilities.
     The basis sets used are aug-cc-pVDZ (aDZ), aug-cc-pVTZ (aTZ),
     and aug-cc-pVQZ (aQZ). 
     Atom C1 is the carbon atom attached to the sulfur atom and H1 is the 
     hydrogen atom attached to C1. 
     Atomic units used for all polarizabilities.
   }
   \label{table:thiophene-ISApol-pols}
\end{table}

In \tabrff{table:thiophene-ISApol-pols} we present \ISApol localized, isotropic 
polarizabilities for the symmetry-distinct atoms in the thiophene molecule
computed in three basis sets.
The dipole--dipole polarizabilities (i.e. rank 1) are already reasonably well converged
in the aug-cc-pVDZ basis, with the exception of the sulfur atom which needs the
larger aug-cc-pVTZ basis.
The quadrupole--quadrupole (rank 2) polarizabilities on the carbon and hydrogen atoms
are converged in the aug-cc-pVTZ basis but the aug-cc-pVQZ basis is needed for the
sulfur atom.
At rank 3, the octopole--octopole polarizabilities on the carbon atoms
seem to be
approaching convergence in the aug-cc-pVQZ basis, but the sulfur atom is far from
convergence. 
The negative octopole--octopole terms on the hydrogen atoms seem to be a 
result of the lack of sufficient higher angular terms on these atoms,
and of the absence of dipole--quadrupole and quadrupole--octopole
polarizabilities in this rather drastic approximation. 
In the aug-cc-pVQZ basis there is only one negative term present on the H1 atom.
Compare these results to those from the \cDF approach shown in 
\tabrff{table:thiophene-cDF-pols}. 
The \ISApol algorithm is clearly the more systematic of the two with the 
AIM local polarizabilities converged or approaching convergence at all ranks.

Dispersion models are obtained from the \ISApolL polarization models computed
at imaginary frequency and recombined using
methods\cite{WilliamsS03}${}^,$\cite[\S4.3.4]{Stone:book:13} 
implemented in the {\sc Casimir} module that forms part of the \CamCASP{}
suite of programs.
While we can compute both anisotropic and isotropic dispersion models, the isotropic
models are easier to analyse and use, so we will focus on these only.

\begin{figure*}
  \includegraphics[width=0.9\textwidth]{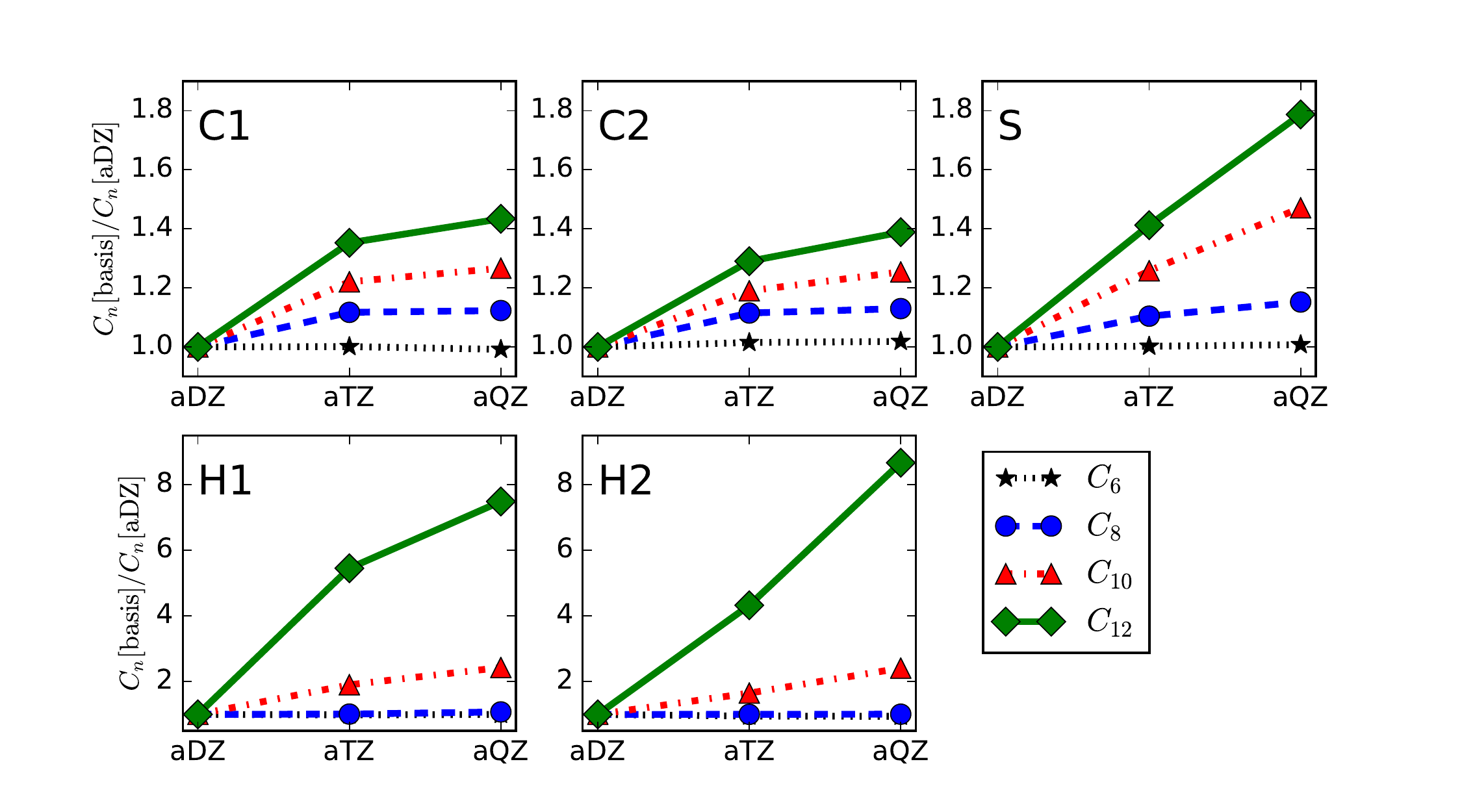}
  \caption[Convergence with basis]{
     Relative dispersion coefficients for the symmetry-distinct sites in thiophene
     computed using the localized, isotropic \ISApol polarizabilities using the
     aug-cc-pVDZ (aDZ), aug-cc-pVTZ (aTZ) and aug-cc-pVQZ (aQZ) basis sets. 
     The dispersion coefficients are relative to the values computed in the
     aug-cc-pVDZ basis. 
  \label{fig:ISApol-Cn-basis-conv}
  }
\end{figure*}

In \figrff{fig:ISApol-Cn-basis-conv} we examine the convergence of the distributed
dispersion models with basis set. 
As the dispersion coefficients span many orders of magnitude, we have instead plotted
the ratio $\C{aa}{n}[\text{basis}]/\C{aa}{n}[\text{aDZ}]$ as a function of
basis set used. This allows us to readily determine how the dispersion coefficients
vary with increasing basis size. 
In the case of the two carbon atoms, the \Cn{6} and \Cn{8} terms have converged
in the aug-cc-pVTZ basis, and the \Cn{10} and \Cn{12} terms nearly so in the 
aug-cc-pVQZ basis. 
For the two hydrogen atoms the \Cn{6} and \Cn{8} terms are converged in the 
aug-cc-pVDZ basis, but the \Cn{10} and \Cn{12} terms are less settled with basis set.
This is probably the result of deficiencies in the higher angular part of the 
hydrogen basis sets, but this needs to be verified.
In any case, the higher ranking dispersion terms do not make a significant 
contribution to the dispersion energy, and have even been fully omitted in
some of our earlier models \cite{MisquittaS08b,MisquittaWSP08}.
However the same cannot be said for the sulfur atom which is expected to
make an important contribution to the dispersion energy due to its large
polarizability: here while the \Cn{6} term is well converged even in the 
aug-cc-pVDZ basis, the \Cn{8} term is only just stabilizing in the aug-cc-pVQZ basis,
and neither \Cn{10} nor \Cn{12} is even close to stabilizing in the
largest basis set used. 
This may be either an artifact of the \ISApol algorithm, or a genuine shortcoming of 
the standard basis sets.
Further and more systematic tests on a wider range of systems will be needed
to determine the cause of this apparent non-convergence.

\begin{table}[ht]
   \newcolumntype{d}{D{.}{.}{6}}
   \begin{tabular}{ldddd}
     \toprule
       Site..Site  
               &\multicolumn{1}{c}{\Cn{6}}
                           &\multicolumn{1}{c}{\Cn{8}}
                                       &\multicolumn{1}{c}{\Cn{10}}
                                                  &\multicolumn{1}{c}{\Cn{12}}\\
     \midrule
     \multicolumn{5}{l}{Pyridine}\\
       C1..C1  &  1.249(1) &  3.141(2) &  1.529(4) &  4.258(5) \\
       C2..C2  &  3.643(1) &  6.525(2) &  3.711(4) &  8.643(5) \\
       C3..C3  &  2.246(1) &  5.555(2) &  2.097(4) &  5.496(5) \\
        N..N   &  3.206(1) &  6.735(2) &  2.609(4) &  6.196(5) \\
       H1..H1  &  3.533(0) &  3.407(1) &  4.384(2) &  4.795(3) \\
       H2..H2  &  1.802(0) &  1.758(1) &  1.921(2) &  1.880(3) \\
       H3..H3  &  1.689(0) &  2.306(1) &  3.392(2) &  4.173(3) \\
     \midrule                                         
     \multicolumn{5}{l}{Water}\\                      
        O..O   &  2.434(1) &  4.899(2) &  1.252(4) &  2.384(5) \\
        H..H   &  0.783(0) &  4.357(0) &  9.061(1) &  7.714(2) \\
     \midrule                                         
     \multicolumn{5}{l}{Methane}\\                    
        C..C   &  3.184(1) &  9.161(2) &  3.771(4) &  1.092(6) \\
        H..H   &  2.105(0) &  2.132(1) &  3.938(2) &  4.638(3) \\
     \midrule                                         
     \multicolumn{5}{l}{Thiophene}\\                    
       C1..C1  & 2.259(1)  & 5.414(2)  & 2.465(4)  & 6.726(5)  \\
       C2..C2  & 3.759(1)  & 8.759(2)  & 4.254(4)  & 1.194(6)  \\
       S..S    & 1.082(2)  & 3.895(3)  & 2.096(5)  & 7.904(6)  \\
       H1..H1  & 2.096(0)  & 2.453(1)  & 2.864(2)  & 3.021(2)  \\
       H2..H2  & 1.268(0)  & 1.630(1)  & 2.518(2)  & 3.207(3)  \\
     \bottomrule
   \end{tabular}
   \caption{
     Localized, isotropic diagonal dispersion coefficients for the symmetry-distinct
     sites in the pyridine, water, methane, and thiophene dimers
     computed with the \ISApolL model.
     The off-diagonal terms, including those between water and methane
     are provided in the S.I.
     These results were computed using the d-aug-cc-pVTZ basis 
     with the exception of the thiophine molecule for which we report
     results computed in the aug-cc-pVQZ basis.
     Due to the large range of numbers involved, the data are provided in
     a compact exponential notation with the power of $10$ indicated in 
     parenthesis. That is $x.y(n) = x.y \times 10^n$.
     Atomic units used for all dispersion coefficients.
   \label{table:ISApol-L-Cn}
   }
\end{table}

In \tabrff{table:ISApol-L-Cn} we report the \ISApolL isotropic dispersion coefficients
for the symmetry-distinct sites in the water, methane, pyridine and 
thiophene molecules. Only the diagonal, that is same-site, terms are reported:
the complete dispersion models for these molecules and also those for the
methane..water complex are given in the S.I.
Notice that while the dispersion coefficients for the carbon atoms in these
molecules are of similar magnitude, they nevertheless vary considerably in 
accordance with what might be expected from the variations in the local 
chemical environment.
For example, the C1 atom in pyridine and the C1 atom in thiophene both have
smaller dispersion coefficients than the other carbon atoms in the
molecules, which should be expected as these atoms are bonded directly to the
more electronegative N and S atoms in the respective molecules.
Likewise, while the dispersion terms on the hydrogen atoms are similar, 
those on the hydrogen atom in water are substantially smaller due to the 
large electronegativity of the oxygen atom in the water molecule. 
The ability of the \ISApolL models to provide dispersion terms from 
\Cn{6} to \Cn{12} which respond to the chemical environment
of the atoms in the molecule could be used to develop more detailed and 
comprehensive models for the dispersion energy, but more extensive
data sets will be needed for a full analysis.

\subsection{Assessing the models using SAPT(DFT)}
\label{sec:assessment}

The ultimate test of any dispersion model is how well it is able to 
match the reference dispersion energies. 
Here, as with the polarization models, there is the issue of damping, without
which meaningful comparisons can only be made at large intermolecular separations
where the damping is negligible. However such a comparison is not useful from
the practical point of view as we are usually interested in the performance of the
models at energetically important configuration, that is, in the region of the
energy minimum.
Consequently we do need to address the issue of damping, but as this is not the
focus of this paper, we will limit the present discussion to the familiar 
Tang--Toennies \cite{TangT84} damping functions:
\begin{align}
  f_n(x) &= 1 - e^{-x} \sum_{k=0}^{n} \frac{x^k}{k!},
  \label{eq:TT}
\end{align}
where the order $n$ corresponds with the rank in the dispersion
expansion \cite{Stone:book:13} and $x$ is a function of the site--site 
distance and the damping coefficient. 
The damping models we have used differ in the definition of $x$ as follows:
\begin{itemize}
  \item Ionization potential (IP) damping \cite{MisquittaS08b}:
    \begin{align}
      x_{ab} &= \left( \sqrt{2 I_A} + \sqrt{2 I_B} \right) r_{ab} = \bb{AB} r_{ab},
      \label{eq:beta-IP}
    \end{align}
    where $I_A$ and $I_B$ are the vertical ionization energies, in a.u., of the two 
    interacting molecules. This is the simplest of the damping models with 
    one damping parameter for all pairs of sites $(a,b)$ between the 
    interacting molecules $A$ and $B$.
  \item The Slater damping from Van Vleet \etal \cite{VanVleetMSS15}. Here the
    damping parameter is dependent on the pairs of interacting atoms and is 
    given by 
    \begin{align}
      x_{ab} &= \bb{ab} r_{ab} - 
                  \frac{\bb{ab}(2\bb{ab} r_{ab} + 3)}
                       {\bb{ab}^2 r_{ab}^2 + 3 \bb{ab} r_{ab} + 3},
      \label{eq:beta-Slater}
    \end{align}
    where the parameter $\bb{ab}$ is now dependent on the sites and is defined
    as $\bb{ab} = \sqrt{\bb{a} \bb{b}}$, where the parameter $\bb{a}$
    is extracted from the ISA shape function $\w{a}$ by
    fitting it to an exponential of the form $K\exp{(-\bb{a} r)}$, and
    $\bb{b}$ likewise.
    \cite{MisquittaSF14,VanVleetMSS15}.
    This damping function is motivated by the form of the overlap of
    two such Slater exponentials\cite{VanVleetMSS15}.
  \item The {\em scaled} ISA damping model is a simplification of the 
    Slater damping model.
    Here we define a scaled parameter $\tbb{a}$ for each site in molecule $A$
    as follows:
    \begin{align}
      \tbb{a} &= s_A \bb{a}, 
      \label{eq:scaled-beta}
    \end{align}
    where $\bb{a}$ is defined above and $s_A$ is the molecule-specific 
    empirical scaling parameter. Next we define $\bb{ab}$ from the combination
    rule
    \begin{align}
      \bb{ab} &= \sqrt{\tbb{a} \tbb{b}},
    \end{align}
    and $x_{ab} = \bb{ab} r_{ab}$.
    In Ref.~\citenum{VanVleetMSS15} the scaling parameter is taken to be a 
    constant $s = 0.84$ independent of the type of molecule, but here we 
    allow the parameter to vary according to the molecule and determine it
    empirically by fitting the model energies to the reference dispersion energies.
\end{itemize}

In the comparisons of the \ISApolL dispersion models that we now discuss, the
reference dispersion energies used in the comparisons have been computed
using SAPT(DFT) and are defined as 
\begin{align}
  \EDISP{2} &= \Edisppol{2} + \Edispexch{2}.
\end{align}
All dispersion models are computed from isotropic \ISApolL polarizabilities,
consequently we should expect errors for systems with a strong anisotropy.
In all cases the isotropic \ISApolL dispersion models contain even terms from
\Cn{6} to \Cn{12} on all atoms.

\begin{figure}
  \includegraphics[width=0.5\textwidth]{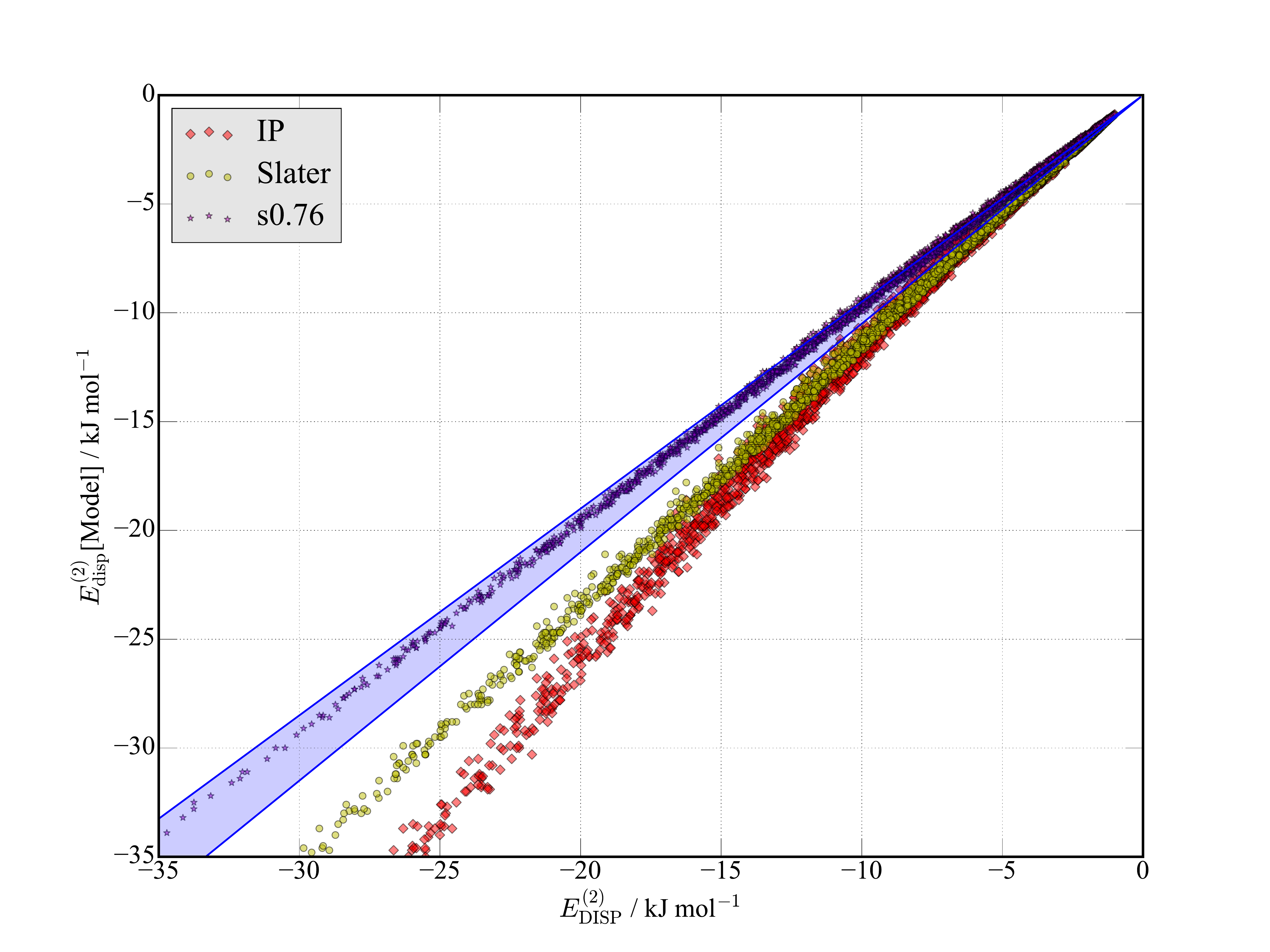}
  \caption[Methane dimer dispersion energies]{
    Dispersion energies for the methane dimer in a variety of configurations.
    Reference energies are computed using SAPT(DFT) as described in the text.
    The \ISApol dispersion models are all isotropic and are damped with various damping
    models: 
    `IP' refers to the Tang--Toennies damping with a single damping parameter
      determined using the molecular ionization potentials, 
    `Slater' refers to the damping model from the Slater-FF model with 
      exponents determined using the ISA, and
    `s0.76' refers to the Tang--Toennies damping with atom-pair-dependent 
      damping parameters determined using the ISA and scaled by $0.76$ as
      described in the text.
    The light blue bar represents $\pm 5$\% errors compared with the reference
    dispersion energies.
  \label{fig:methane2-disp}
  }
\end{figure}

In \figrff{fig:methane2-disp} we display dispersion energies for the methane dimer
in more than 2600 dimer configurations. 
Because the methane molecule has high symmetry and indeed is nearly 
spherical, we should expect the dispersion energy of this system to be 
well approximated by an isotropic dispersion model. 
This is indeed the case, and we see nearly perfect correlation of the 
\ISApolL dispersion energies with the {\em scaled damping model} with the 
reference energies. In this case a scaling parameter of $0.76$ was determined.
On the other hand, the IP damping model which we have recommended in the past
does not provide sufficient damping, and nor does the Slater model,
though it is better.

\begin{figure}
  \includegraphics[width=0.5\textwidth]{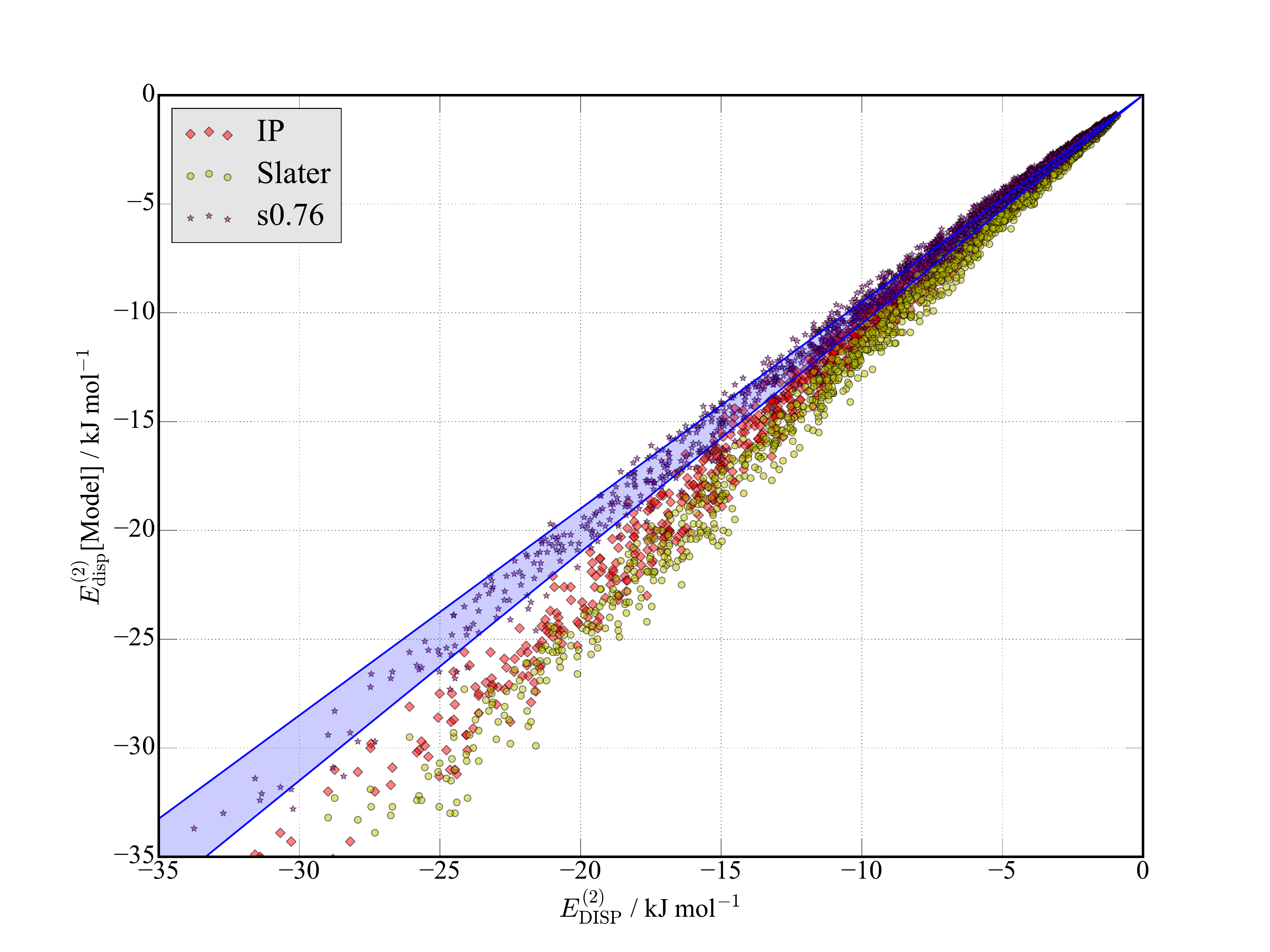}
  \caption[Water dimer dispersion energies]{
    Dispersion energies for the water dimer in a variety of configurations.
    Reference energies are computed using SAPT(DFT) as described in the text.
    The \ISApol dispersion models are all isotropic and are damped with various damping
    models which are described in the caption to \figrff{fig:methane2-disp}.
  \label{fig:water2-disp}
  }
\end{figure}

\begin{figure}
  \includegraphics[width=0.5\textwidth]{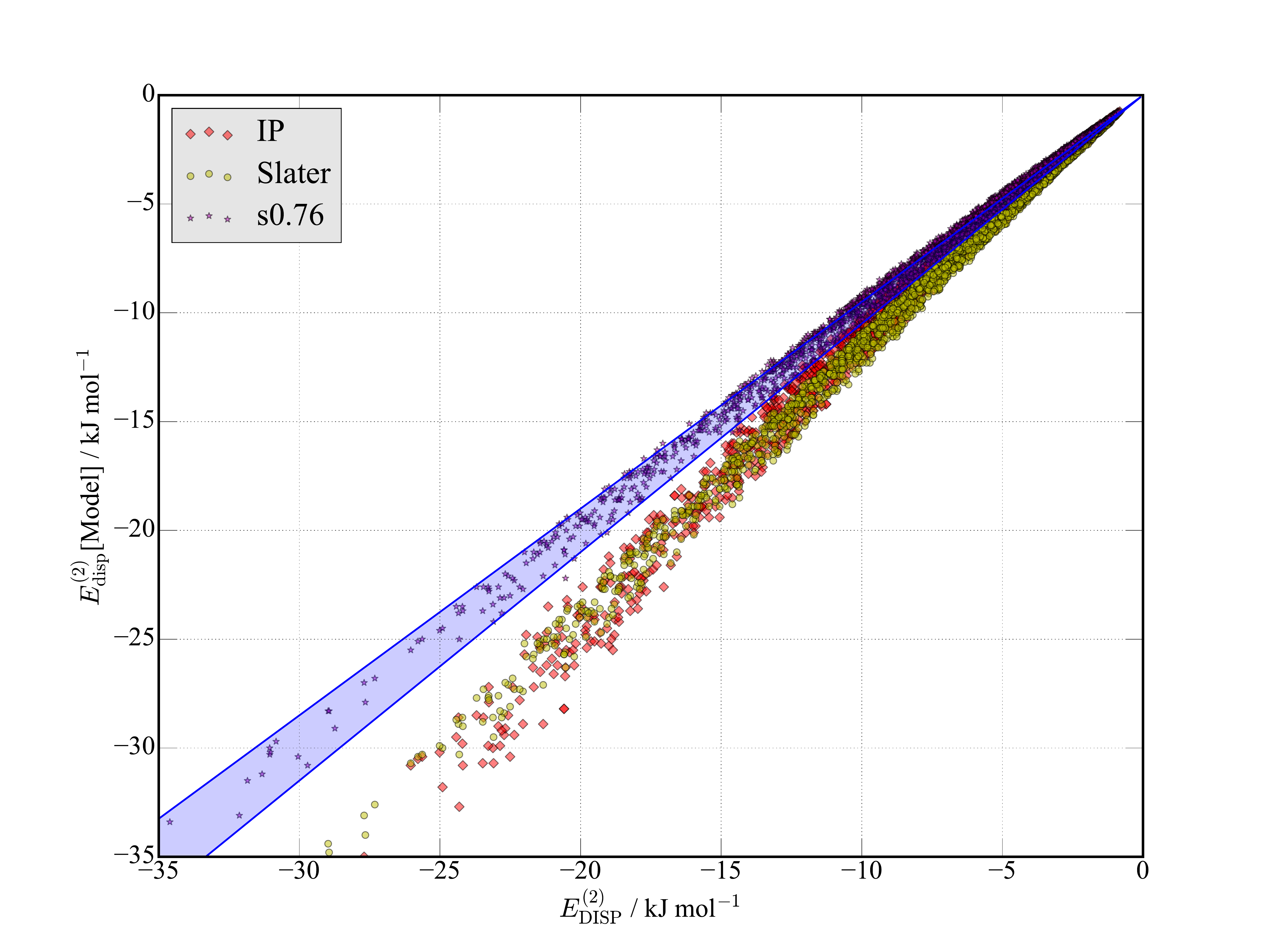}
  \caption[Water dimer dispersion energies]{
    Dispersion energies for the methane..water dimer in a variety of configurations.
    Reference energies are computed using SAPT(DFT) as described in the text.
    The \ISApol dispersion models are all isotropic and are damped with various damping
    models which are described in the caption to \figrff{fig:methane2-disp}.
  \label{fig:methane-water-disp}
  }
\end{figure}

\figrff{fig:water2-disp} shows data for the water dimer in more than
2000 dimer configurations. 
Water is a more anisotropic system than methane, and we cannot expect the isotropic
models to behave as well for water dimer as for methane dimer.
Once again both the IP and Slater damping models result in underdamping, though
not as severely as for methane dimer. 
The scaled damping model with a scaling factor of $0.76$ fares far better, resulting
in dispersion energies for most of the dimers within $\pm 5$\% from the reference
energies.
In \figrff{fig:methane-water-disp} we have displayed dispersion energies for the
mixed methane\vdw water system. The picture is the same, with the scaled damping 
model correlating very well with the reference energies.

\begin{figure}
  \includegraphics[width=0.5\textwidth]{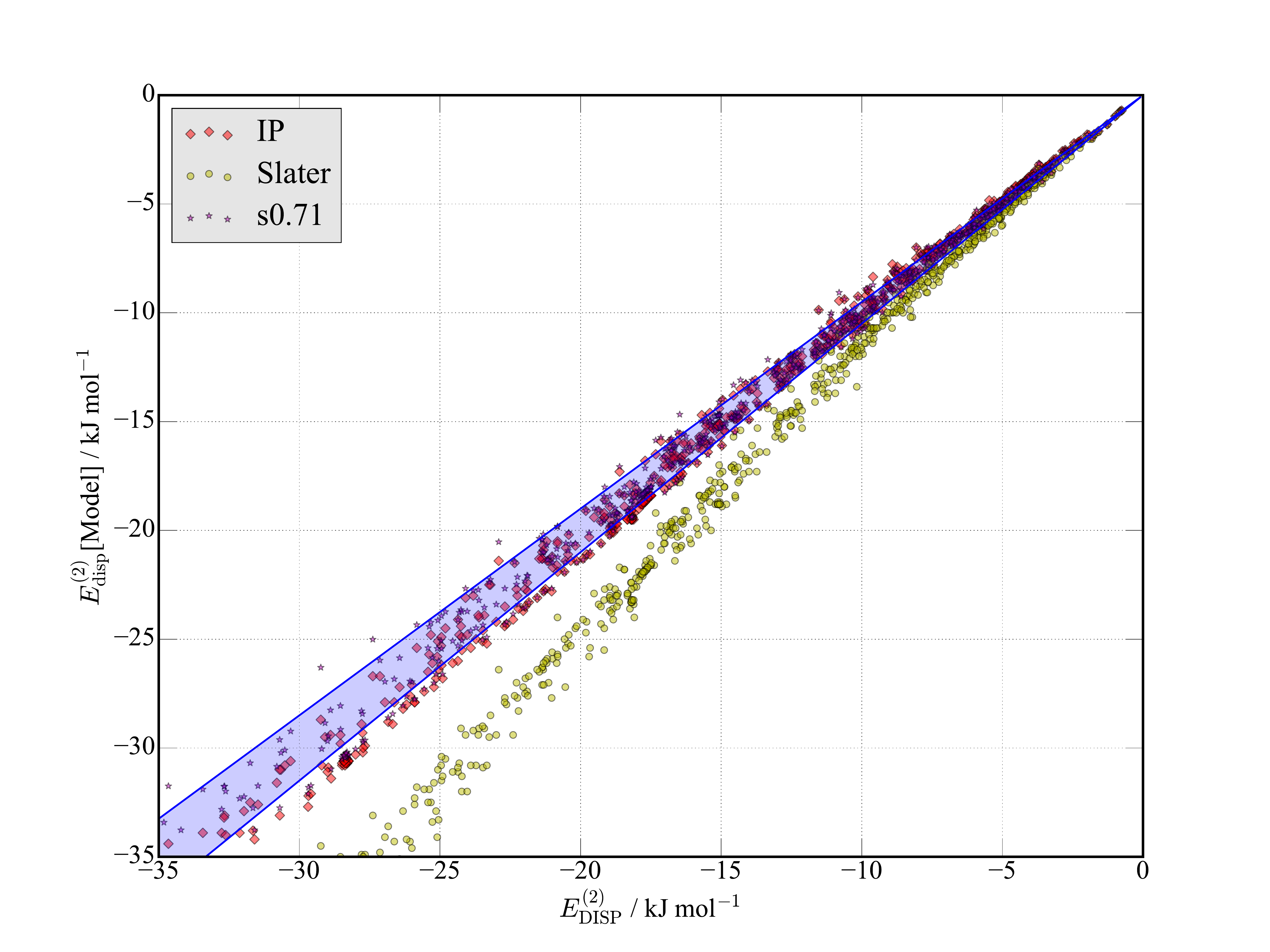}
  \caption[pyridine dimer dispersion energies]{
    Dispersion energies for the pyridine dimer in a variety of configurations.
    Reference energies are computed using SAPT(DFT) as described in the text.
    The \ISApol dispersion models are all isotropic and are damped with various damping
    models which are described in the caption to \figrff{fig:methane2-disp}.
  \label{fig:pyridine2-disp}
  }
\end{figure}

In \figrff{fig:pyridine2-disp} we display dispersion energies for the 
pyridine dimer in over 700 configurations taken from data sets 1 and 2
from Ref.~\citenum{MisquittaS16}.
The pyridine molecule is the most anisotropic one we have considered in this
paper and we may therefore expect to see a relatively large scatter in the 
model dispersion energies. This is indeed the case: while the scaled damping 
dispersion model still results in the best dispersion energies, these now 
deviate from the reference energies by slightly more than $5$\%. 
The scaling parameter has been determined to be $0.71$ which is smaller than the
values obtained for the water and methane systems, and considerably smaller than
the value of $0.84$ recommended by Van Vleet \etal \cite{VanVleetMSS15}
Part of the reason for this is that the AIM densities for the pyridine 
molecule are themselves strongly anisotropic due to the $\pi$-electron 
density of the molecule, but the parameters $\bb{a}$ used in 
\eqrff{eq:scaled-beta} are obtained from the isotropic shape functions
and therefore the correct AIM density decay is not obtained. Instead
the anisotropic AIM densities $\R{a}$ should be used, and we are
currently investigating this possibility. 
Curiously, for this system the IP damping model is quite similar to the
scaled damping, but the Slater damping model once again under-damps.

\subsection{Convergence with rank}
\label{sec:conv-with-rank}

\begin{figure}
  \includegraphics[width=0.5\textwidth]{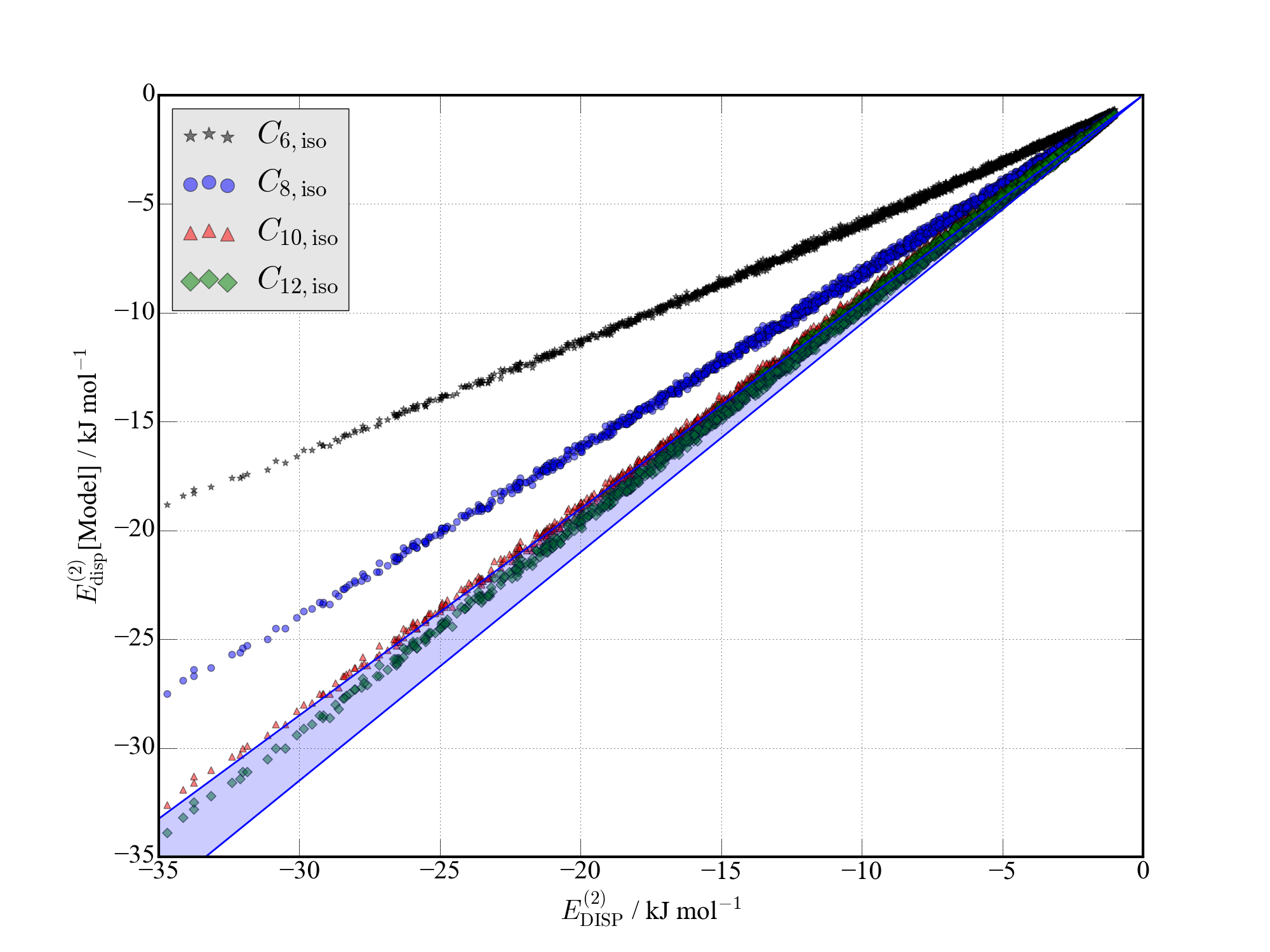}
  \caption[methane dimer dispersion energies at various ranks]{
    Dispersion energies for the methane dimer from \ISApol isotropic
    dispersion models at various maximum ranks. 
    All models are damped using the scaled Tang--Toennies damping with
    scaling $0.76$. 
  \label{fig:methane2-disp-var-rank}
  }
\end{figure}

Although it is reasonably well known that the dispersion expansion should 
include terms beyond \Cn{6}, it is perhaps not as well appreciated just 
how many terms are required for this expansion to converge (when appropriately
damped). We have explored this issue in a previous paper \cite{MisquittaS08b},
where we concluded that models including terms to at least \Cn{10} were needed
to achieve sufficiently good agreement with SAPT(DFT). 
In \figrff{fig:methane2-disp-var-rank} we present even more extensive data
for the methane dimer which clearly demonstrates that the \Cn{6}-only models
commonly used in simple force-fields, and indeed in many dispersion corrections
to density-functional theory, severely underestimate the 
dispersion energy from SAPT(DFT). For this dimer, we need to include terms
to \Cn{10} before we begin to agree with the reference energies to within 5\%. 

\subsection{Combination rules}
\label{sec:comb-rules}

Dispersion models in common intermolecular interaction models are usually
constructed to satisfy combination rules, usually through a
constrained fitting process (see for example
Ref.~\citenum{McDanielS13}).
This has the advantage of
greatly reducing the number of parameters in the model, and the most
commonly used {\em geometric mean}
combination rule has good justification from theory, although 
the actual dispersion coefficients
may not satisfy a combination rule accurately.

The geometric mean combination rule
defines the mixed site \C{ab}{n} dispersion coefficients as follows:
\begin{align}
  \C{ab}{n} = \sqrt{\C{aa}{n}\, \C{bb}{n}},
  \label{eq:geom-mean}
\end{align}
where $\C{aa}{n}$ and $\C{bb}{n}$ are the same-site coefficients.
This combination rule may be derived for the $n=6$ terms \cite{MavroyannisS62} from the
exact expression for the isotropic \C{ab}{6} coefficient:
\begin{align}
  \C{ab}{6} &= \frac{3}{\pi} \int_{0}^{\infty} 
                   \bar{\alpha}^{a}(i v) \bar{\alpha}^{b}(i v) dv,
\end{align}
by using the single-pole approximation to the isotropic frequency-dependent
polarizabilities
\begin{align}
  \bar{\alpha}(i v) &= \bar{\alpha}(0) \frac{v_0^2}{v^2 + v_0^2},
  \label{eq:single-pole}
\end{align}
where $v_0$ is the pole. We additionally have to assume that
the poles for the two sites $a$ and $b$ are similar, that is, $v_0^a \approx v_0^b$.
This is identical to the Uns\"{o}ld average energy approximation \cite{Stone:book:13}.
The advantages of this combination rule are apparent: for a system of $N$ interacting
sites, only $\order{N}$ dispersion coefficients would be needed, rather than the
$\order{N^2}$ needed without such a rule. 

\begin{figure}
  \includegraphics[width=0.5\textwidth]{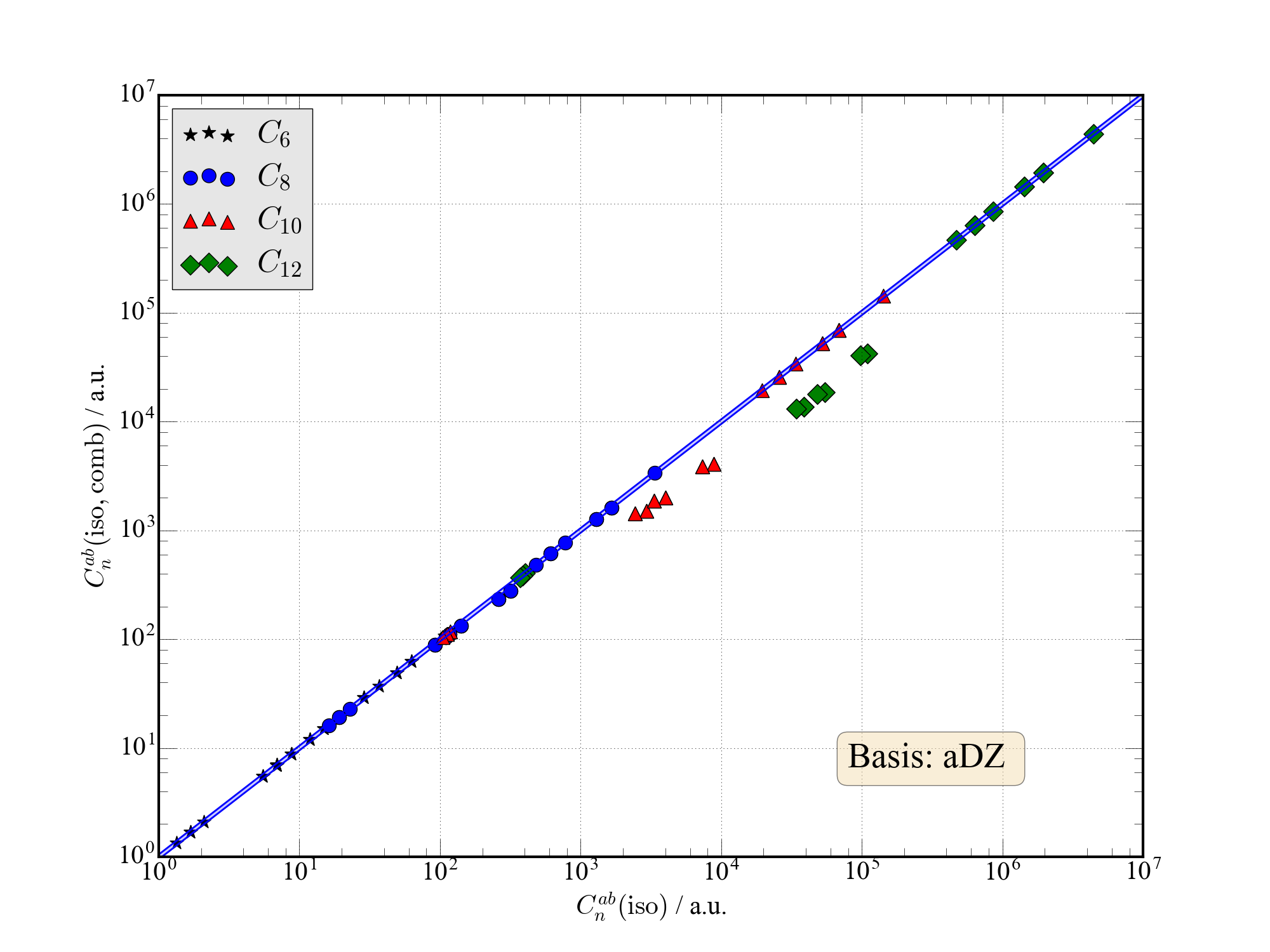}
  \includegraphics[width=0.5\textwidth]{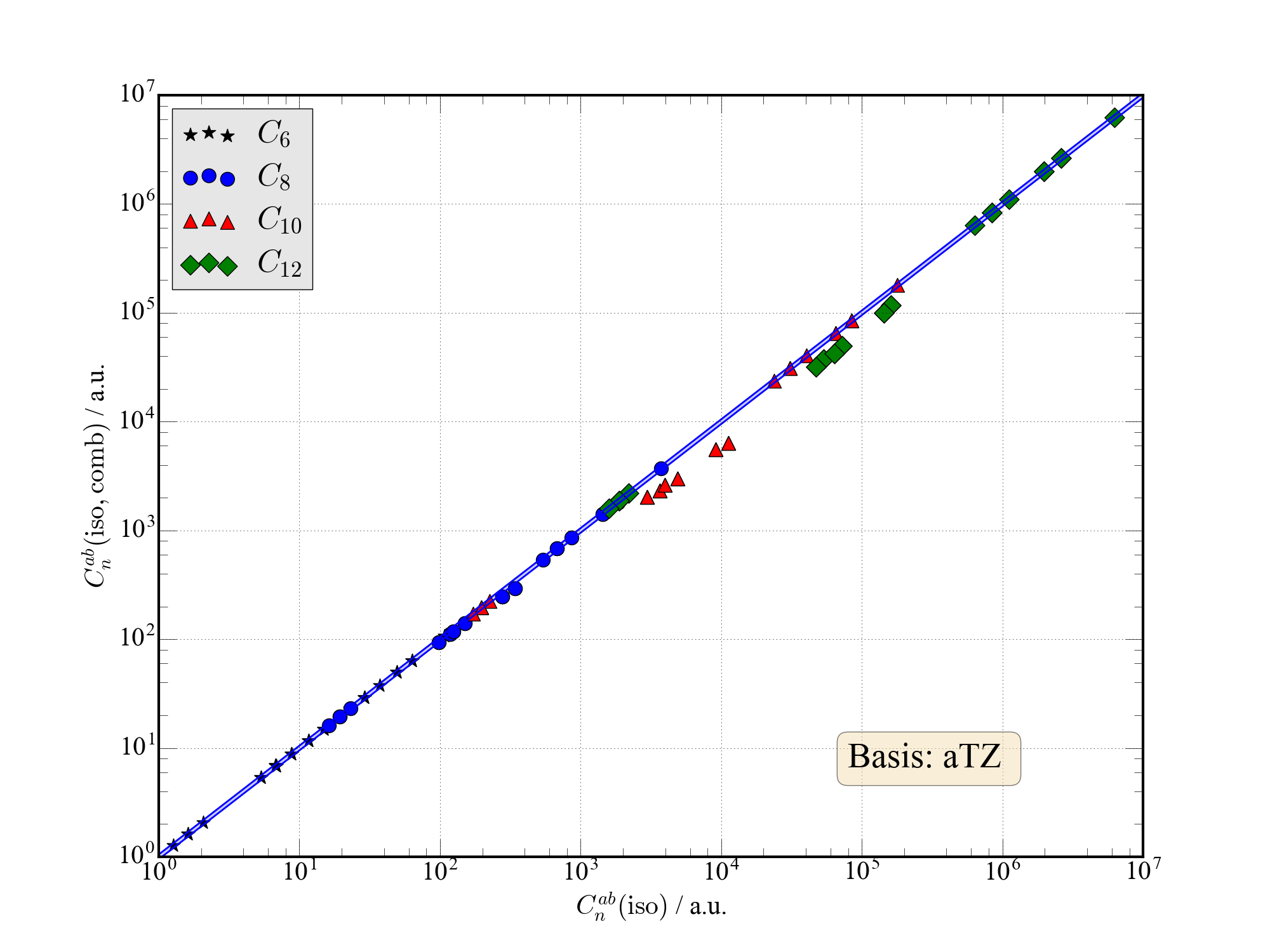}
  \includegraphics[width=0.5\textwidth]{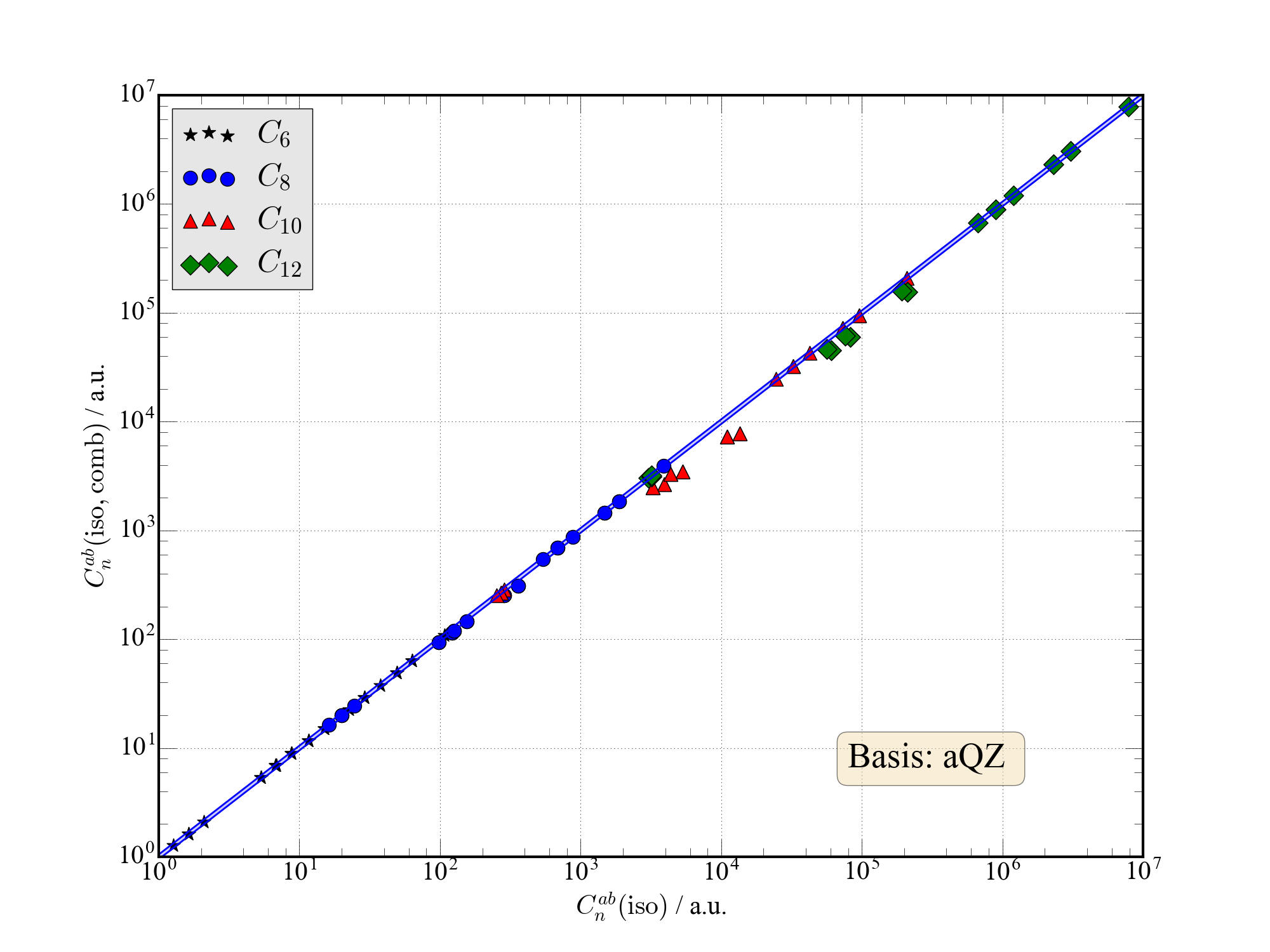}
  \caption[Combination rules]{
    Comparison of \ISApol dispersion coefficients for thiophene 
    against those obtained using the geometric mean combination rule. 
    The three panels show how the combination rules are satisfied as a function
    of the basis set used to obtain the \ISApol isotropic dispersion models.
    In all cases the points off the diagonal line are associated with 
    the hydrogen atoms in thiophene. 
  \label{fig:comb-rules-thiophene2}
  }
\end{figure}

Do the \ISApol dispersion models satisfy the geometric mean combination rule?
Once again this question is a complex one if we account for the angular variation
of the dispersion parameters, so here we will restrict this discussion to the
isotropic dispersion models only. 
In \figrff{fig:comb-rules-thiophene2} we plot the dispersion coefficients for 
the thiophene molecule computed using the geometric mean combination rule
against reference \ISApolL isotropic dispersion coefficients.
This is done for the aug-cc-pV$n$Z, $n = \text{D,T,Q}$ basis sets.
It can be seen that the \ISApolL models satisfy the combination rule very well
for $n=6,8,10,12$, that is for all ranks of the dispersion coefficients considered
in this paper. In all cases, the terms that are most in error are those
involving at least one of the hydrogen atoms, but these errors are reduced as the
basis set gets larger, echoing the trend to more well-defined polarizabilities
seen in \tabrff{table:thiophene-ISApol-pols}.

This property of the dispersion models derived from \ISApolL polarizabilities
seems to hold for a variety of systems, though less well for those containing
a larger fraction of hydrogen atoms. This is remarkable given that the 
combination rules are never imposed, and there is no reason to expect the 
single-pole approximation to hold, or indeed for the poles on different atoms
to be similar. 
Further work is needed to analyse exactly 
why this is the case, and if and when it breaks down,
but this property of the \ISApolL models, if generally applicable,
will be a very useful feature
for the development of models of more diverse interactions.

\section{Analysis \& Outlook}
\label{sec:analysis}

We have described and implemented the \ISApol algorithm for computing distributed
frequency-dependent polarizabilities and dispersion coefficients for
molecular systems. 
This algorithm is based on a basis-space implementation \cite{MisquittaSF14}
of the iterated stockholder atoms (ISA) algorithm of Lillestolen
and Wheatley \cite{LillestolenW08}.
We have described a simpler and more versatile implementation of the 
BS-ISA algorithm and have implemented this algorithm in a developer's
version of \CamCASP{6.0}.
This new algorithm allows for higher accuracies in the ISA solution and 
in the resulting distributed properties. 
Additionally the algorithm has a computational cost that scales linearly
with the system size. 

The \ISApol algorithm results in non-local distributed polarizabilities
which can be localized to result in approximate atomic polarizabilities
using schemes we have discussed and demonstrated.
The resulting models have many of the desired properties discussed
in the Introduction. The most important of these are:
\begin{itemize}
  \item Systematic convergence of the \ISApol non-local polarizabilities
    as a function of rank. This model has been demonstrated to converge 
    more systematically than the constrained density fitting, \cDF, 
    model we have previously proposed \cite{MisquittaS06}, 
    and also the related \SRLO algorithm from Rob \& Szalewicz \cite{RobS13a}. 
  \item The localized \ISApol polarizabilities (\ISApolL) are well defined and are 
    usually positive definite where local models can give a good
    account of what are inherently non-local effects. 
    In other words, for systems with relatively short electron correlation lengths,
    the \ISApolL models are appropriate and systematic and lead to reasonably accurate 
    polarization energies.
  \item We have demonstrated that the \ISApolL polarizabilities converge systematically
    with basis set and appear to have a well-defined basis set limit.
    The systematic behaviour of these distributed polarizabilities should make
    it possible to extrapolate the polarizabilities of the atoms-in-the-molecule
    (AIMs) to the complete basis set limit.
    This was not possible with the \WSM models \cite{MisquittaS08a,MisquittaSP08}
    built from \cDF non-local polarizabilities as has been illustrated in the
    Introduction.
  \item Dispersion models constructed from the \ISApolL frequency-dependent 
    polarizabilities are well defined and, when suitably damped,
    show exceptionally good reproduction of the\break SAPT(DFT) dispersion energies
    for a variety of anisotropic systems. 
  \item Damping of the dispersion models is achieved using the Tang--Toennies 
    functions with atom-specific damping parameters derived using the BS-ISA
    algorithm. A single scaling parameter is used as described by Van Vleet \etal
    \cite{VanVleetMSS15}, though we have allowed the scaling parameter to vary with the
    molecule.
  \item The isotropic dispersion coefficients from the \ISApolL algorithm
    have been shown to satisfy the geometric mean combination rule that is 
    used in many empirical models for the dispersion energy, but is not
    imposed at any stage in developing the localized \ISApol polarizabilities.
    This is the case for terms from \Cn{6} to \Cn{12} and the accuracy
    of the combination rule improves with increase in the basis set used
    for the \ISApol calculation. 
\end{itemize}
These properties alone make the \ISApol and associated localized \ISApolL models
promising candidates for developing detailed and accurate polarization and 
dispersion models for intermolecular interactions. 
At present, these methods are limited to closed-shell molecules, but this is 
to a large extent a limitation of the implementation in the \CamCASP{6.0} program.

Amongst the issues that we have not yet resolved adequately are the 
determination of the damping of the polarization and dispersion models,
and the problem of the anisotropy of the dispersion models.
The polarization damping question has been raised by one of us elsewhere
\cite{Misquitta13a} but it needs to be re-visited in context of the
\ISApol models for which the damping needed is clearly different from models
derived from the \cDF polarizabilities (see \secrff{sec:conv-rank}).
The damping models introduced by Van Vleet \etal \cite{VanVleetMSS15}
are definitely promising. In particular, we have shown that the scaled ISA 
damping model can result in dispersion energies that agree with the 
reference SAPT(DFT) total dispersion energy, \EDISP{2}, to 5\% or better.
In fact, for the methane dimer the agreement is much better than 5\%,
and also substantially better than that achieved by a recently proposed anisotropic
LoProp-based dispersion model \cite{HarczukNJVA17}.
However there remains the question of how this can be improved and it seems like 
there are a few issues that need to be investigated:
\begin{itemize}
  \item {\em Anisotropy in the damping}: Perhaps the damping coefficients need
    to be extracted from the ISA AIM densities \R{a} rather than from the 
    ISA shape functions \w{a} as we do currently. This would have the consequence
    of making the damping parameters anisotropic and these may be more
    appropriate at modelling interactions involving sites that are 
    themselves strongly anisotropic. This would be the case for the oxygen atom
    in water and for the carbon atoms in a $\pi$-conjugated system. 
  \item {\em Anisotropy in the dispersion coefficients}: The dispersion models 
    derived from the \ISApolL polarizabilities include anisotropy, but we have,
    as yet, focused only on the isotropic parts of these models. 
    This has been done mainly for computational reasons: most simulation codes 
    accept only isotropic dispersion models, and the anisotropic models tend
    to be very complex. Recently, Van Vleet \etal \cite{VanVleetMS18} have
    demonstrated how the inclusion of atomic anisotropy can result in a 
    rather significant improvement in the model energies, but this approach
    is empirical in the sense that the anisotropy parameters are determined
    by fitting to reference SAPT(DFT) dispersion energies. 
    We need a way to develop practical models in a non-empirical manner.
\end{itemize}

We have not investigated the transferability of the \ISApol polarizabilities
as these are not the fundamental AIM polarizabilities, but are effective 
atomic polarizabilities after through-space polarization in the
Applequist sense \cite{ApplequistCF72,Stone:book:13} has been taken into account.
It should however be possible to derive the `bare' AIM polarizabilities
from those computed from \ISApol and this is something we are currently 
exploring.
Finally, the fundamental relation of the \ISApol models with the underlying
ISA decomposition may eventually lead to the development of approximations
that allow the models to be mapped onto the properties of the ISA AIM densities. 
If possible, this would significantly increase our ability to easily construct
polarization models for complex molecular system, especially those too large
for routine linear-response calculations in a large enough basis set.
This too is something we are currently exploring. 

\section{Acknowledgements}
We dedicate this article to the memory of
Dr J{\'a}nos {\'A}ngy{\'a}n for a friendship and for many scientific discussions,
one of which led to the \ISApol method.
AJM thanks Queen Mary University of London for support and the
Thomas Young Centre for a stimulating environment, and 
also the Universit{\'e} de Lorraine for a
visiting professorship during which part of this work was completed.
We also thank Dr Rory A. J. Gilmore for assistance in
calculating the SAPT(DFT) reference energies for the water..water,
methane..methane, and water..methane complexes.
We thank Dr Toon Verstraelen for helpful comments on the manuscript.

\section{Additional information}
All developments have been implemented in a developer's version of the
\CamCASP{6.0} \cite{CamCASP} program which may be obtained from the
authors on request. \CamCASP{} has been interfaced to the \Dalton (2006 through to 2015), 
\NWChem, \Gamess, and \PsiF programs.
The supplementary information (SI) contains additional data from 
the systems we have investigated but not included in this paper.

\setlength\bibsep{2pt}
\bibliography{BIBLIOGRAPHY}

\end{document}


\title[Distributed polarizabilities from the ISA]
{\ISApol: Distributed polarizabilities and dispersion models from a
basis-space implementation of the iterated stockholder atoms procedure : SI}

\author{Alston J. Misquitta}
\affiliation{School of Physics and Astronomy, 
and the Thomas Young Centre for Theory and Simulation of Materials,
Queen Mary, University of London,
London E1 4NS, U.K.}
\author{Anthony J. Stone}
\affiliation{University Chemical Laboratory, Lensfield Road,
Cambridge, CB2 1EW, U.K.}

\date{\today}

\begin{abstract}
Supplementary Information for the \ISApol paper.
\end{abstract}

\maketitle



\section{Figures and tables from the \SRLO and \cDF methods}

These figures and tables are the analogues of the \ISApol data presented in
the main body of the paper. These are provided here so as to facilitate comparisons
with these methods and the \ISApol approach.

\subsection{Convergence of distributed polarizabilities with basis}
The data for the \ISApolL and \cDFL approaches are in the main paper. Here 
we present the data for the \SRLOL method.

\begin{table}
   \newcolumntype{d}{D{.}{.}{2}}
   \begin{tabular}{llddd}
     \toprule
       Site  & $l$ &\multicolumn{1}{r}{\text{aDZ}}
                               &\multicolumn{1}{r}{\text{aTZ}}
                                         &\multicolumn{1}{r}{\text{aQZ}}\\
     \midrule
       C1    &  1  &     7.26  &   7.21  &    6.96  \\
             &  2  &    29.93  &  32.09  &   19.33  \\
             &  3  &  -351.37  &  30.40  & 1270.26  \\
     \midrule
       C2    &  1  &    10.64  &  10.81  &   11.10  \\
             &  2  &    43.92  &  58.37  &   44.72  \\
             &  3  &  -596.91  &-475.20  &  966.75  \\
     \midrule
       S     &  1  &    16.62  &  16.88  &   16.91  \\
             &  2  &    80.06  &  96.53  &  108.49  \\
             &  3  &  -571.82  &-906.44  &  -19.89  \\
     \midrule
       H1    &  1  &     2.24  &   2.25  &    2.36  \\
             &  2  &    -0.25  &  -2.88  &    6.44  \\
             &  3  &   -95.06  & -31.28  & -193.66  \\
     \midrule
       H2    &  1  &     1.56  &   1.47  &    1.37  \\
             &  2  &    -0.28  &  -3.86  &    7.73  \\
             &  3  &   -61.37  & -18.68  & -214.82  \\
     \bottomrule
   \end{tabular}
   \caption{
     Localized, isotropic polarizabilities for the symmetry-distinct
     sites in the thiophene molecule computed with the \SRLOL model, that
     is using \SRLO non-local polarizabilities localized using the
     \WSM algorithm.
     The basis sets used are aug-cc-pVDZ (aDZ), aug-cc-pVTZ (aTZ),
     and aug-cc-pVQZ (aQZ). 
     Atom C1 is the carbon atom attached to the sulfur atom and H1 is the 
     hydrogen atom attached to C1. 
     Atomic units used for all polarizabilities.
   }
   \label{table:thiophene-SRLO-pols}
\end{table}

\pagebreak

\subsection{Convergence of distributed dispersion coefficients with thiophene}

\begin{figure*}
  \includegraphics[width=0.9\textwidth]{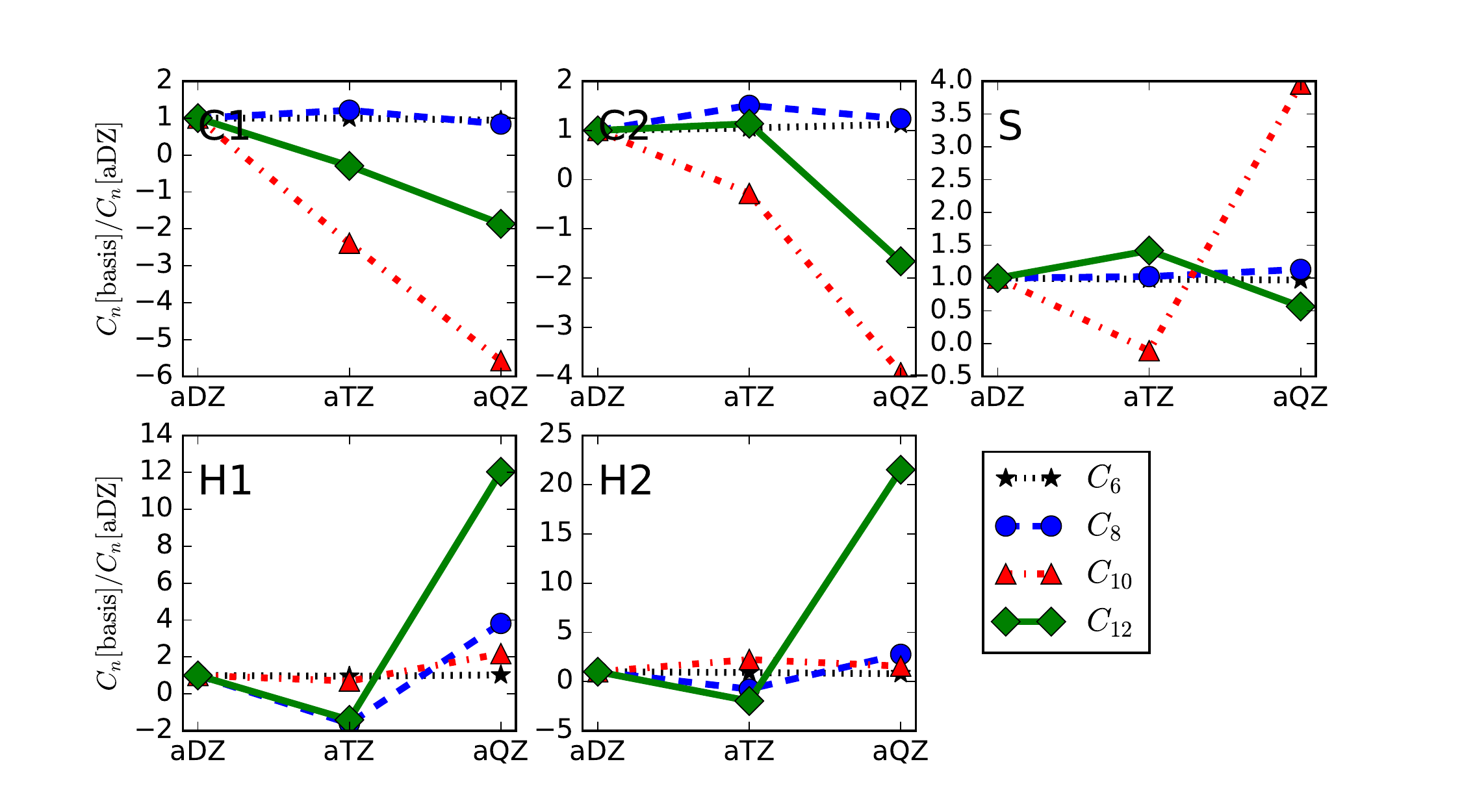}
  \caption[Convergence with basis]{
     Relative dispersion coefficients for the symmetry-distinct sites in thiophene
     computed using the localized, isotropic \cDF polarizabilities using the
     aug-cc-pVDZ (aDZ), aug-cc-pVTZ (aTZ) and aug-cc-pVQZ (aQZ) basis sets. 
     The dispersion coefficients are relative to the values computed in the
     aug-cc-pVDZ basis. 
  \label{fig:cDF-Cn-basis-conv}
  }
\end{figure*}

\begin{figure*}
  \includegraphics[width=0.9\textwidth]{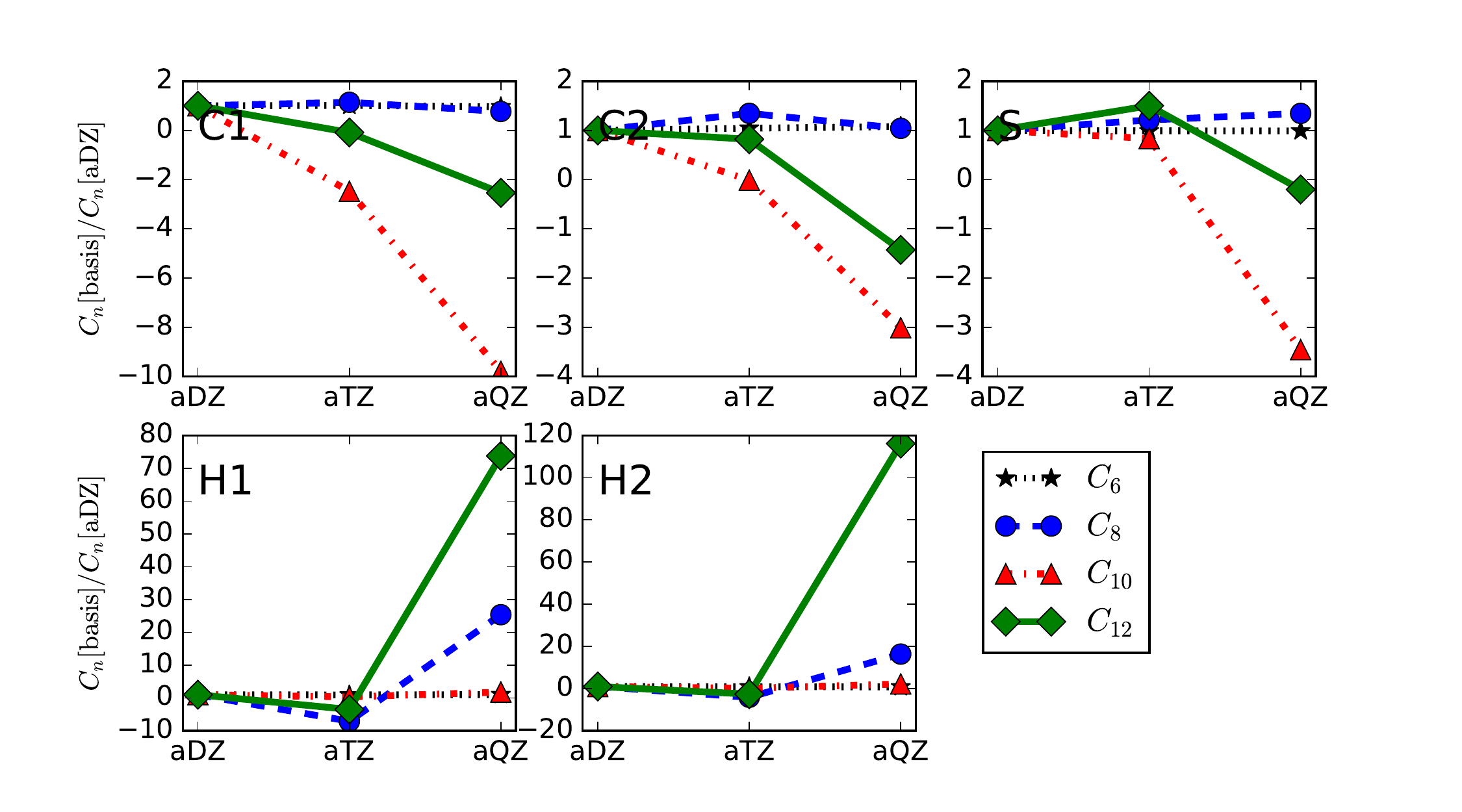}
  \caption[Convergence with basis]{
     Relative dispersion coefficients for the symmetry-distinct sites in thiophene
     computed using the localized, isotropic \SRLO polarizabilities using the
     aug-cc-pVDZ (aDZ), aug-cc-pVTZ (aTZ) and aug-cc-pVQZ (aQZ) basis sets. 
     The dispersion coefficients are relative to the values computed in the
     aug-cc-pVDZ basis. 
  \label{fig:SRLO-Cn-basis-conv}
  }
\end{figure*}

\pagebreak

\section{Geometries \& local-axis definitions}
The molecular geometries used for the molecules studied in this paper are provided here
along with the local axis systems used to define the localized polarizabilities.
The axis systems are defined in a way suitable for use in the \Orient program
and the reader is refered to the documentation of that program for further details.

\subsection{Methane}
\begin{lstlisting}
Molecule CH4
  I.P. 0.4634 a.u.
  ! Vibrationally averaged geom.
  C   6.0    0.0000000000    0.0000000000    0.0000000000 Type C 
  H1  1.0    0.0000000000    2.0770357780    0.0000000000 Type H
  H2  1.0    1.6958926074   -0.6923452530    0.9791240615 Type H
  H3  1.0   -1.6958926074   -0.6923452530    0.9791240615 Type H
  H4  1.0    0.0000000000   -0.6923452530   -1.9582481041 Type H  
End
\end{lstlisting}
\begin{lstlisting}
Axes
  C  z from C to H1       x from H3 to H2
  H1 z from C to H1       x from H3 to H2
  H2 z from C to H2       x from H4 to H1
  H3 z from C to H3       x from H1 to H4
  H4 z from C to H4       x from H2 to H3
End
\end{lstlisting}

\subsection{Water}
\begin{lstlisting}
Molecule H2O
  I.P. 0.4638 a.u.
  Units Bohr
  ! Vibrationally averaged geom.
  O   8.0    0.0000000000    0.0000000000    0.0000000000 Type O
  H1  1.0   -1.4536519600    0.0000000000   -1.1216873200 Type H
  H2  1.0    1.4536519600    0.0000000000   -1.1216873200 Type H
End
\end{lstlisting}
\begin{lstlisting}
! Molecule in the xz-plane
Axes
  O  z between H1 and H2  x from H1 to H2
  H1 z from  O to H1      x from H1 to H2
  H2 z from  O to H2      x from H2 to H1
End
\end{lstlisting}

\subsection{Pyridine}
\begin{lstlisting}
Molecule pyridine
    ! Optimzed with PBE0/cc-pVTZ Gaussian03
    ! C2v symmetry
    IP   0.3488  a.u.
    Units Angstrom
    H1 1.0     -2.050322    1.274414    0.000000
    H2 1.0     -2.147113   -1.203259    0.000000
    H3 1.0      0.000000   -2.487558    0.000000
    H4 1.0      2.147113   -1.203259    0.000000
    H5 1.0      2.050322    1.274414    0.000000
    N  7.0      0.000000    1.382844    0.000000
    C1 6.0     -1.134410    0.690452    0.000000
    C2 6.0     -1.190513   -0.695795    0.000000
    C3 6.0      0.000000   -1.403912    0.000000
    C4 6.0      1.190513   -0.695795    0.000000
    C5 6.0      1.134410    0.690452    0.000000
End
\end{lstlisting}

\begin{lstlisting}
Axes
  N   x from C3 to N   y from N  to C1
  C1  x from C1 to H1  y from C1 to C2
  H1  x from C1 to H1  y from C1 to C2
  C5  x from C5 to H5  y from C5 to C4
  H5  x from C5 to H5  y from C5 to C4
  C2  x from C2 to H2  y from C2 to C3
  H2  x from C2 to H2  y from C2 to C3
  C4  x from C4 to H4  y from C4 to C3
  H4  x from C4 to H4  y from C4 to C3
  C3  x from C3 to H3  y from C3 to C4
  H3  x from C3 to H3  y from C3 to C4
End
\end{lstlisting}

\subsection{Thiophene}
\begin{lstlisting}
Molecule thiophene
  ! Optimized with PBE0/cc-pVTZ  NWChem
  ! C2v symmetry
  I.P. 0.326 a.u.
  Units Angstrom
  C1  6       0.70902646     1.27833645     0.00000000  Type C1
  C2  6      -0.70902646     1.27833645     0.00000000  Type C1
  C3  6       1.23164616     0.02041947     0.00000000  Type C2
  C4  6      -1.23164616     0.02041947     0.00000000  Type C2
  S  16       0.00000000    -1.16811386     0.00000000  Type S
  H1  1       1.31493664     2.17399255     0.00000000  Type H1
  H2  1      -1.31493664     2.17399255     0.00000000  Type H1
  H3  1       2.26927521    -0.27351467     0.00000000  Type H2
  H4  1      -2.26927521    -0.27351467     0.00000000  Type H2
End
\end{lstlisting}

\begin{lstlisting}
! x axis of all atoms is along the bond, pointing out.
! For S, we define the y-axis first, and the x-axis is define by
! Schmidt orthogonalization to point from between C4..C3 to S.
Axes
  S  y from C4 to C3   x from C3 to S
  C1 x from C1 to H1   y from C1 to C3
  H1 x from C1 to H1   y from C1 to C3
  C2 x from C2 to H2   y from C2 to C4
  H2 x from C2 to H2   y from C2 to C4
  C3 x from C3 to H3   y from C3 to S
  H3 x from C3 to H3   y from C3 to S
  C4 x from C4 to H4   y from C4 to S
  H4 x from C4 to H4   y from C4 to S
End
\end{lstlisting}

\section{Dispersion models}
These dispersion models are in a format suitable for use with the \Orient program.
The damping models used are already presented and these too are in a format 
that can be used with \Orient.
Atomic units are used. 

\subsection{Methane dimer}
\begin{lstlisting}
!  Localisation settings for CH4 
!  Axes file:       CH4.axes  
!  Pol file format: NEW  
!  Limit:           3  
!  WSM-Limit:       3  
!  H-Limit:         3  
!  Isotropic?:      True  
!  Model file:      CH4.pdef  
!  Pol Cutoff:      0.0001  
!  Loc algorithm:   LW  
!  Weight:          3  
!  Weight coeff:    1e-05  
!  SVD threshold:   0.0  
!  NoRefine?:       False  
!                       
! Dispersion coefficients for CH4 (hartree bohr^n)
  C  C               C6             C7             C8             C9             C10            C11            C12
    00   00   0   31.83934         0.0          916.1360         0.0          37705.16         0.0          1091999.
  End
  H  C               C6             C7             C8             C9             C10            C11            C12
    00   00   0   8.144934         0.0          158.1968         0.0          4955.234         0.0          75282.47
  End
  H  H               C6             C7             C8             C9             C10            C11            C12
    00   00   0   2.104675         0.0          21.31545         0.0          393.8321         0.0          4638.035
  End
\end{lstlisting}

\begin{lstlisting}
IP-damping:
===========
   ! We will use a damping model based on the Tang--Toennies incomplete Gamma function
   ! with damping parameter beta = 1.9254 a.u.
   ! based on CH4 IP of 0.4634 a.u.
 
   Dispersion damping factor 1.9254
 

Slater-damping:
===============
!  Damping model :  SLATER  
!  Algorithm =  avg  
!  Scale     =  1.000000  
!  
      C    C  
         Dispersion DAMPING TT-SLATER   1.8193 
      End 
      C    H  
         Dispersion DAMPING TT-SLATER   2.1845 
      End 
      H    C  
         Dispersion DAMPING TT-SLATER   2.1845 
      End 
      H    H  
         Dispersion DAMPING TT-SLATER   2.5497 
      End 


Scaled-ISA-damping:
==================
!  Damping model :  TT  
!  Algorithm =  avg  
!  Scale     =  0.760000  
!  
      C    C  
         Dispersion damping factor      1.3827 
      End 
      C    H  
         Dispersion damping factor      1.6602 
      End 
      H    C  
         Dispersion damping factor      1.6602 
      End 
      H    H  
         Dispersion damping factor      1.9378 
      End 
\end{lstlisting}

\subsection{Water dimer}
\begin{lstlisting}
!  Localisation settings for H2O 
!  Axes file:       H2O.axes  
!  Pol file format: NEW  
!  Limit:           3  
!  WSM-Limit:       3  
!  H-Limit:         3  
!  Isotropic?:      True  
!  Model file:      H2O.pdef  
!  Pol Cutoff:      0.0001  
!  Loc algorithm:   LW  
!  Weight:          3  
!  Weight coeff:    1e-05  
!  SVD threshold:   0.0  
!  NoRefine?:       False  
!                       
! Dispersion coefficients for H2O (hartree bohr^n)
  O  O               C6             C7             C8             C9             C10            C11            C12
    00   00   0   24.34089         0.0          489.9063         0.0          12519.45         0.0          238364.1
  End
  H  O               C6             C7             C8             C9             C10            C11            C12
    00   00   0   4.335086         0.0          55.94859         0.0          1174.193         0.0          13116.46
  End
  H  H               C6             C7             C8             C9             C10            C11            C12
    00   00   0  0.7833591         0.0          4.356823         0.0          90.61106         0.0          771.3764
  End
\end{lstlisting}

\begin{lstlisting}
IP-damping:
===========
   ! We will use a damping model based on the Tang--Toennies incomplete Gamma function
   ! with damping parameter beta = 1.9262 a.u.
   ! based on CH4 IP of 0.4638 a.u.
 
   Dispersion damping factor 1.9262


Slater-damping:
===============
!  Damping model :  SLATER  
!  Algorithm =  avg  
!  Scale     =  1.000000  
!  
      O    O  
         Dispersion DAMPING TT-SLATER   2.3413 
      End 
      O    H  
         Dispersion DAMPING TT-SLATER   2.5014 
      End 
      H    O  
         Dispersion DAMPING TT-SLATER   2.5014 
      End 
      H    H  
         Dispersion DAMPING TT-SLATER   2.6615 
      End 

Scaled-ISA-damping:
===================
!  Damping model :  TT  
!  Algorithm =  avg  
!  Scale     =  0.760000  
!  
      O    O  
         Dispersion damping factor      1.7794 
      End 
      O    H  
         Dispersion damping factor      1.9011 
      End 
      H    O  
         Dispersion damping factor      1.9011 
      End 
      H    H  
         Dispersion damping factor      2.0227 
      End 

\end{lstlisting}

\subsection{Methane..water complex}
\begin{lstlisting}
! CH4..H2O
! ISA-Pol daTZ PBE0/AC L3iso wt3 coeff1e-5
! 
! Dispersion coefficients for CH4 and H2O (hartree bohr^n)
  C  O               C6             C7             C8             C9             C10            C11            C12
    00   00   0   27.74901         0.0          679.9535         0.0          23092.61         0.0          527158.6
  End
  C  HO              C6             C7             C8             C9             C10            C11            C12
    00   00   0   4.989649         0.0          85.63989         0.0          2627.002         0.0          28044.23
  End
  HC  O              C6             C7             C8             C9             C10            C11            C12
    00   00   0   7.042453         0.0          106.9746         0.0          2351.979         0.0          33449.61
  End
  HC  HO             C6             C7             C8             C9             C10            C11            C12
    00   00   0   1.281617         0.0          10.04097         0.0          187.0358         0.0          1851.580
  End
\end{lstlisting}

\begin{lstlisting}
IP-damping:
===========
   ! We will use a damping model based on the Tang--Toennies incomplete Gamma function
   ! with damping parameter beta = 1.9258 a.u.
   ! based on CH4 IP of 0.4634 a.u. and H2O IP of 0.4638 a.u.
 
   Dispersion damping factor 1.9258
 

Slater-damping:
===============
!  Damping model :  SLATER  
!  Algorithm =  avg  
!  Scale     =  1.000000  
!  
      C    O  
         Dispersion DAMPING TT-SLATER   2.0803 
      End 
      C    HO  
         Dispersion DAMPING TT-SLATER   2.2404 
      End 
      HC    O  
         Dispersion DAMPING TT-SLATER   2.4455 
      End 
      HC    HO  
         Dispersion DAMPING TT-SLATER   2.6056 
      End 


Scaled-ISA-damping:
===================
!  Damping model :  TT  
!  Algorithm =  avg  
!  Scale     =  0.760000  
!  
      C    O  
         Dispersion damping factor      1.5810 
      End 
      C    HO  
         Dispersion damping factor      1.7027 
      End 
      HC    O  
         Dispersion damping factor      1.8586 
      End 
      HC    HO  
         Dispersion damping factor      1.9803 
      End 

\end{lstlisting}

\subsection{Pyridine dimer}
\begin{lstlisting}
!  Localisation settings for pyridine 
!  Axes file:       pyridine.axes  
!  Pol file format: NEW  
!  Limit:           3  
!  WSM-Limit:       3  
!  H-Limit:         3  
!  Isotropic?:      True  
!  Model file:      pyridine.pdef  
!  Pol Cutoff:      0.0001  
!  Loc algorithm:   LW  
!  Weight:          3  
!  Weight coeff:    0.001  
!  SVD threshold:   0.0  
!  NoRefine?:       False  
!                       
! Dispersion coefficients for pyridine (hartree bohr^n)
  H1  H1             C6             C7             C8             C9             C10            C11            C12
    00   00   0   3.532866         0.0          34.06937         0.0          438.4128         0.0          4795.142
  End
  H2  H1             C6             C7             C8             C9             C10            C11            C12
    00   00   0   2.512308         0.0          24.42942         0.0          291.5739         0.0          3031.013
  End
  H2  H2             C6             C7             C8             C9             C10            C11            C12
    00   00   0   1.802484         0.0          17.57851         0.0          192.0930         0.0          1880.357
  End
  H3  H1             C6             C7             C8             C9             C10            C11            C12
    00   00   0   2.429523         0.0          28.39730         0.0          387.3553         0.0          4472.532
  End
  H3  H2             C6             C7             C8             C9             C10            C11            C12
    00   00   0   1.744662         0.0          20.41735         0.0          261.8713         0.0          2807.164
  End
  H3  H3             C6             C7             C8             C9             C10            C11            C12
    00   00   0   1.689167         0.0          23.06164         0.0          339.2002         0.0          4172.905
  End
  N  H1              C6             C7             C8             C9             C10            C11            C12
    00   00   0   10.59577         0.0          161.5577         0.0          4633.989         0.0          63777.48
  End
  N  H2              C6             C7             C8             C9             C10            C11            C12
    00   00   0   7.568185         0.0          116.8522         0.0          3249.586         0.0          42793.63
  End
  N  H3              C6             C7             C8             C9             C10            C11            C12
    00   00   0   7.329912         0.0          127.6860         0.0          3618.335         0.0          60441.76
  End
  N  N               C6             C7             C8             C9             C10            C11            C12
    00   00   0   32.05721         0.0          673.5358         0.0          26087.00         0.0          619564.3
  End
  C1  H1             C6             C7             C8             C9             C10            C11            C12
    00   00   0   6.497265         0.0          110.5402         0.0          4062.762         0.0          57787.89
  End
  C1  H2             C6             C7             C8             C9             C10            C11            C12
    00   00   0   4.694808         0.0          80.87772         0.0          2867.138         0.0          39224.61
  End
  C1  H3             C6             C7             C8             C9             C10            C11            C12
    00   00   0   4.555179         0.0          87.48131         0.0          3100.917         0.0          54765.35
  End
  C1  N              C6             C7             C8             C9             C10            C11            C12
    00   00   0   19.84052         0.0          456.3821         0.0          20270.59         0.0          519046.6
  End
  C1  C1             C6             C7             C8             C9             C10            C11            C12
    00   00   0   12.48911         0.0          314.1222         0.0          15292.93         0.0          425844.0
  End
  C2  H1             C6             C7             C8             C9             C10            C11            C12
    00   00   0   11.31604         0.0          155.4416         0.0          6205.428         0.0          87544.46
  End
  C2  H2             C6             C7             C8             C9             C10            C11            C12
    00   00   0   8.041769         0.0          113.1853         0.0          4367.324         0.0          59837.66
  End
  C2  H3             C6             C7             C8             C9             C10            C11            C12
    00   00   0   7.783660         0.0          125.1831         0.0          4689.731         0.0          83183.31
  End
  C2  N              C6             C7             C8             C9             C10            C11            C12
    00   00   0   34.12012         0.0          666.8137         0.0          31210.98         0.0          753850.7
  End
  C2  C1             C6             C7             C8             C9             C10            C11            C12
    00   00   0   20.97147         0.0          455.1061         0.0          23684.95         0.0          609509.2
  End
  C2  C2             C6             C7             C8             C9             C10            C11            C12
    00   00   0   36.42686         0.0          652.5485         0.0          37114.84         0.0          864282.2
  End
  C3  H1             C6             C7             C8             C9             C10            C11            C12
    00   00   0   8.853158         0.0          150.6656         0.0          4169.249         0.0          58730.11
  End
  C3  H2             C6             C7             C8             C9             C10            C11            C12
    00   00   0   6.343089         0.0          109.2201         0.0          2943.922         0.0          39341.48
  End
  C3  H3             C6             C7             C8             C9             C10            C11            C12
    00   00   0   6.145039         0.0          118.0003         0.0          3289.226         0.0          55530.84
  End
  C3  N              C6             C7             C8             C9             C10            C11            C12
    00   00   0   26.81696         0.0          613.2778         0.0          23331.98         0.0          583434.4
  End
  C3  C1             C6             C7             C8             C9             C10            C11            C12
    00   00   0   16.66157         0.0          415.2136         0.0          18098.01         0.0          493814.0
  End
  C3  C2             C6             C7             C8             C9             C10            C11            C12
    00   00   0   28.49454         0.0          610.6677         0.0          27578.47         0.0          719098.0
  End
  C3  C3             C6             C7             C8             C9             C10            C11            C12
    00   00   0   22.45808         0.0          555.5270         0.0          20973.81         0.0          549561.5
  End
\end{lstlisting}

\begin{lstlisting}
IP-damping:
===========
   ! We will use a damping model based on the Tang--Toennies incomplete Gamma function
   ! with damping parameter \beta = 1.67 a.u.
 
   Dispersion damping factor 1.67
 
Slater-damping:
===============
!  Files pyr_beta.dat     pyr_beta.dat   
!  Damping model :  SLATER  
!  Algorithm =  geom  
!  Scale     =  1.000000  
!  
      H1    H1  
         Dispersion DAMPING TT-SLATER   2.3834 
      End 
      H1    H2  
         Dispersion DAMPING TT-SLATER   2.4620 
      End 
      H1    H3  
         Dispersion DAMPING TT-SLATER   2.4181 
      End 
      H1    N  
         Dispersion DAMPING TT-SLATER   2.3021 
      End 
      H1    C1  
         Dispersion DAMPING TT-SLATER   2.3092 
      End 
      H1    C2  
         Dispersion DAMPING TT-SLATER   2.2507 
      End 
      H1    C3  
         Dispersion DAMPING TT-SLATER   2.2929 
      End 
      H2    H1  
         Dispersion DAMPING TT-SLATER   2.4620 
      End 
      H2    H2  
         Dispersion DAMPING TT-SLATER   2.5432 
      End 
      H2    H3  
         Dispersion DAMPING TT-SLATER   2.4979 
      End 
      H2    N  
         Dispersion DAMPING TT-SLATER   2.3780 
      End 
      H2    C1  
         Dispersion DAMPING TT-SLATER   2.3854 
      End 
      H2    C2  
         Dispersion DAMPING TT-SLATER   2.3249 
      End 
      H2    C3  
         Dispersion DAMPING TT-SLATER   2.3685 
      End 
      H3    H1  
         Dispersion DAMPING TT-SLATER   2.4181 
      End 
      H3    H2  
         Dispersion DAMPING TT-SLATER   2.4979 
      End 
      H3    H3  
         Dispersion DAMPING TT-SLATER   2.4534 
      End 
      H3    N  
         Dispersion DAMPING TT-SLATER   2.3356 
      End 
      H3    C1  
         Dispersion DAMPING TT-SLATER   2.3429 
      End 
      H3    C2  
         Dispersion DAMPING TT-SLATER   2.2835 
      End 
      H3    C3  
         Dispersion DAMPING TT-SLATER   2.3263 
      End 
      N    H1  
         Dispersion DAMPING TT-SLATER   2.3021 
      End 
      N    H2  
         Dispersion DAMPING TT-SLATER   2.3780 
      End 
      N    H3  
         Dispersion DAMPING TT-SLATER   2.3356 
      End 
      N    N  
         Dispersion DAMPING TT-SLATER   2.2235 
      End 
      N    C1  
         Dispersion DAMPING TT-SLATER   2.2304 
      End 
      N    C2  
         Dispersion DAMPING TT-SLATER   2.1739 
      End 
      N    C3  
         Dispersion DAMPING TT-SLATER   2.2146 
      End 
      C1    H1  
         Dispersion DAMPING TT-SLATER   2.3092 
      End 
      C1    H2  
         Dispersion DAMPING TT-SLATER   2.3854 
      End 
      C1    H3  
         Dispersion DAMPING TT-SLATER   2.3429 
      End 
      C1    N  
         Dispersion DAMPING TT-SLATER   2.2304 
      End 
      C1    C1  
         Dispersion DAMPING TT-SLATER   2.2373 
      End 
      C1    C2  
         Dispersion DAMPING TT-SLATER   2.1806 
      End 
      C1    C3  
         Dispersion DAMPING TT-SLATER   2.2215 
      End 
      C2    H1  
         Dispersion DAMPING TT-SLATER   2.2507 
      End 
      C2    H2  
         Dispersion DAMPING TT-SLATER   2.3249 
      End 
      C2    H3  
         Dispersion DAMPING TT-SLATER   2.2835 
      End 
      C2    N  
         Dispersion DAMPING TT-SLATER   2.1739 
      End 
      C2    C1  
         Dispersion DAMPING TT-SLATER   2.1806 
      End 
      C2    C2  
         Dispersion DAMPING TT-SLATER   2.1254 
      End 
      C2    C3  
         Dispersion DAMPING TT-SLATER   2.1652 
      End 
      C3    H1  
         Dispersion DAMPING TT-SLATER   2.2929 
      End 
      C3    H2  
         Dispersion DAMPING TT-SLATER   2.3685 
      End 
      C3    H3  
         Dispersion DAMPING TT-SLATER   2.3263 
      End 
      C3    N  
         Dispersion DAMPING TT-SLATER   2.2146 
      End 
      C3    C1  
         Dispersion DAMPING TT-SLATER   2.2215 
      End 
      C3    C2  
         Dispersion DAMPING TT-SLATER   2.1652 
      End 
      C3    C3  
         Dispersion DAMPING TT-SLATER   2.2058 
      End 

Scaled-ISA-damping:
===================
!  Files pyr_beta.dat     pyr_beta.dat   
!  Damping model :  TT  
!  Algorithm =  avg  
!  Scale     =  0.710000  
!  
      H1    H1  
         Dispersion damping factor      1.6922 
      End 
      H1    H2  
         Dispersion damping factor      1.7489 
      End 
      H1    H3  
         Dispersion damping factor      1.7171 
      End 
      H1    N  
         Dispersion damping factor      1.6354 
      End 
      H1    C1  
         Dispersion damping factor      1.6403 
      End 
      H1    C2  
         Dispersion damping factor      1.6006 
      End 
      H1    C3  
         Dispersion damping factor      1.6292 
      End 
      H2    H1  
         Dispersion damping factor      1.7489 
      End 
      H2    H2  
         Dispersion damping factor      1.8057 
      End 
      H2    H3  
         Dispersion damping factor      1.7738 
      End 
      H2    N  
         Dispersion damping factor      1.6922 
      End 
      H2    C1  
         Dispersion damping factor      1.6971 
      End 
      H2    C2  
         Dispersion damping factor      1.6574 
      End 
      H2    C3  
         Dispersion damping factor      1.6859 
      End 
      H3    H1  
         Dispersion damping factor      1.7171 
      End 
      H3    H2  
         Dispersion damping factor      1.7738 
      End 
      H3    H3  
         Dispersion damping factor      1.7419 
      End 
      H3    N  
         Dispersion damping factor      1.6603 
      End 
      H3    C1  
         Dispersion damping factor      1.6652 
      End 
      H3    C2  
         Dispersion damping factor      1.6255 
      End 
      H3    C3  
         Dispersion damping factor      1.6540 
      End 
      N    H1  
         Dispersion damping factor      1.6354 
      End 
      N    H2  
         Dispersion damping factor      1.6922 
      End 
      N    H3  
         Dispersion damping factor      1.6603 
      End 
      N    N  
         Dispersion damping factor      1.5787 
      End 
      N    C1  
         Dispersion damping factor      1.5836 
      End 
      N    C2  
         Dispersion damping factor      1.5439 
      End 
      N    C3  
         Dispersion damping factor      1.5724 
      End 
      C1    H1  
         Dispersion damping factor      1.6403 
      End 
      C1    H2  
         Dispersion damping factor      1.6971 
      End 
      C1    H3  
         Dispersion damping factor      1.6652 
      End 
      C1    N  
         Dispersion damping factor      1.5836 
      End 
      C1    C1  
         Dispersion damping factor      1.5885 
      End 
      C1    C2  
         Dispersion damping factor      1.5488 
      End 
      C1    C3  
         Dispersion damping factor      1.5773 
      End 
      C2    H1  
         Dispersion damping factor      1.6006 
      End 
      C2    H2  
         Dispersion damping factor      1.6574 
      End 
      C2    H3  
         Dispersion damping factor      1.6255 
      End 
      C2    N  
         Dispersion damping factor      1.5439 
      End 
      C2    C1  
         Dispersion damping factor      1.5488 
      End 
      C2    C2  
         Dispersion damping factor      1.5090 
      End 
      C2    C3  
         Dispersion damping factor      1.5376 
      End 
      C3    H1  
         Dispersion damping factor      1.6292 
      End 
      C3    H2  
         Dispersion damping factor      1.6859 
      End 
      C3    H3  
         Dispersion damping factor      1.6540 
      End 
      C3    N  
         Dispersion damping factor      1.5724 
      End 
      C3    C1  
         Dispersion damping factor      1.5773 
      End 
      C3    C2  
         Dispersion damping factor      1.5376 
      End 
      C3    C3  
         Dispersion damping factor      1.5661 
      End 
\end{lstlisting}

\subsection{Thiophene dimer}
Here we include the \ISApolL, \cDFL and \SRLOL dispersion models for 
thiophene as calculated using the aug-cc-pVQZ basis. Details of the
electronic structure methods used are given in the main paper.

\subsubsection{\ISApolL}
\begin{lstlisting}
!  Localisation settings for thiophene_aQZ 
!  Axes file:       thiophene.axes  
!  Pol file format: NEW  
!  Limit:           3  
!  WSM-Limit:       3  
!  H-Limit:         3  
!  Isotropic?:      True  
!  Model file:      thiophene_aQZ.pdef  
!  Pol Cutoff:      0.0001  
!  Loc algorithm:   LW  
!  Weight:          3  
!  Weight coeff:    0.001  
!  SVD threshold:   0.0  
!  NoRefine?:       False  
!                       
! Dispersion coefficients for thiophene (hartree bohr^n)
  C1  C1             C6             C7             C8             C9             C10            C11            C12
    00   00   0   22.59204         0.0          541.4023         0.0          24646.88         0.0          672582.9
  End
  C2  C1             C6             C7             C8             C9             C10            C11            C12
    00   00   0   29.03714         0.0          687.0662         0.0          32403.42         0.0          897766.8
  End
  C2  C2             C6             C7             C8             C9             C10            C11            C12
    00   00   0   37.58707         0.0          875.9451         0.0          42540.05         0.0          1193955.
  End
  S  C1              C6             C7             C8             C9             C10            C11            C12
    00   00   0   49.22080         0.0          1473.384         0.0          72983.38         0.0          2301622.
  End
  S  C2              C6             C7             C8             C9             C10            C11            C12
    00   00   0   63.69818         0.0          1889.864         0.0          95400.87         0.0          3059320.
  End
  S  S               C6             C7             C8             C9             C10            C11            C12
    00   00   0   108.2377         0.0          3895.374         0.0          209604.8         0.0          7904383.
  End
  H1  C1             C6             C7             C8             C9             C10            C11            C12
    00   00   0   6.851789         0.0          121.6726         0.0          3940.330         0.0          60890.06
  End
  H1  C2             C6             C7             C8             C9             C10            C11            C12
    00   00   0   8.828863         0.0          155.3583         0.0          5308.818         0.0          83207.57
  End
  H1  S              C6             C7             C8             C9             C10            C11            C12
    00   00   0   15.03045         0.0          358.8927         0.0          13501.33         0.0          211381.8
  End
  H1  H1             C6             C7             C8             C9             C10            C11            C12
    00   00   0   2.095591         0.0          24.53272         0.0          286.4169         0.0          3020.599
  End
  H2  C1             C6             C7             C8             C9             C10            C11            C12
    00   00   0   5.311494         0.0          98.34884         0.0          3262.526         0.0          56131.48
  End
  H2  C2             C6             C7             C8             C9             C10            C11            C12
    00   00   0   6.796336         0.0          125.2834         0.0          4380.473         0.0          76131.47
  End
  H2  S              C6             C7             C8             C9             C10            C11            C12
    00   00   0   11.58775         0.0          285.1373         0.0          10972.35         0.0          192610.5
  End
  H2  H1             C6             C7             C8             C9             C10            C11            C12
    00   00   0   1.624049         0.0          20.02837         0.0          272.0111         0.0          3178.793
  End
  H2  H2             C6             C7             C8             C9             C10            C11            C12
    00   00   0   1.268042         0.0          16.30085         0.0          251.8105         0.0          3207.063
  End
\end{lstlisting}

\subsubsection{\cDFL}
\begin{lstlisting}
!  Localisation settings for thiophene_aQZ 
!  Axes file:       thiophene.axes  
!  Pol file format: NEW  
!  Limit:           3  
!  WSM-Limit:       3  
!  H-Limit:         3  
!  Isotropic?:      True  
!  Model file:      None  
!  Pol Cutoff:      0.0001  
!  Loc algorithm:   LW  
!  Weight:          3  
!  Weight coeff:    0.001  
!  SVD threshold:   0.0  
!  NoRefine?:       False  
!                       
! Dispersion coefficients for thiophene (hartree bohr^n)
  C1  C1             C6             C7             C8             C9             C10            C11            C12
    00   00   0   20.50553         0.0          479.4246         0.0          34183.51         0.0          1065041.
  End
  C2  C1             C6             C7             C8             C9             C10            C11            C12
    00   00   0   29.23557         0.0          693.4308         0.0          38292.06         0.0          1178352.
  End
  C2  C2             C6             C7             C8             C9             C10            C11            C12
    00   00   0   41.98772         0.0          1020.457         0.0          40334.55         0.0          1141084.
  End
  S  C1              C6             C7             C8             C9             C10            C11            C12
    00   00   0   45.46070         0.0          1286.247         0.0          43662.00         0.0          1517694.
  End
  S  C2              C6             C7             C8             C9             C10            C11            C12
    00   00   0   65.42397         0.0          1902.285         0.0          41613.05         0.0          986667.8
  End
  S  S               C6             C7             C8             C9             C10            C11            C12
    00   00   0   102.6069         0.0          3477.760         0.0          38202.11         0.0         -855392.1
  End
  H1  C1             C6             C7             C8             C9             C10            C11            C12
    00   00   0   6.826780         0.0          132.7218         0.0          3881.727         0.0          55630.13
  End
  H1  C2             C6             C7             C8             C9             C10            C11            C12
    00   00   0   9.802184         0.0          197.7211         0.0          2228.688         0.0         -13256.68
  End
  H1  S              C6             C7             C8             C9             C10            C11            C12
    00   00   0   15.44321         0.0          383.8911         0.0         -1810.201         0.0         -263123.0
  End
  H1  H1             C6             C7             C8             C9             C10            C11            C12
    00   00   0   2.340054         0.0          36.77741         0.0         -1042.155         0.0         -34702.66
  End
  H2  C1             C6             C7             C8             C9             C10            C11            C12
    00   00   0   4.783793         0.0          140.2342         0.0          3375.391         0.0          125484.5
  End
  H2  C2             C6             C7             C8             C9             C10            C11            C12
    00   00   0   6.799728         0.0          203.9525         0.0          2533.488         0.0          42317.28
  End
  H2  S              C6             C7             C8             C9             C10            C11            C12
    00   00   0   10.71628         0.0          369.1795         0.0          1028.557         0.0         -251292.6
  End
  H2  H1             C6             C7             C8             C9             C10            C11            C12
    00   00   0   1.635039         0.0          41.22609         0.0         -637.2519         0.0         -39899.79
  End
  H2  H2             C6             C7             C8             C9             C10            C11            C12
    00   00   0   1.160629         0.0          39.86654         0.0         -181.0389         0.0         -43833.92
  End
\end{lstlisting}

\subsubsection{\SRLOL}
\begin{lstlisting}
!  Localisation settings for thiophene_aQZ 
!  Axes file:       thiophene.axes  
!  Pol file format: NEW  
!  Limit:           3  
!  WSM-Limit:       3  
!  H-Limit:         3  
!  Isotropic?:      True  
!  Model file:      None  
!  Pol Cutoff:      0.0001  
!  Loc algorithm:   LW  
!  Weight:          3  
!  Weight coeff:    0.001  
!  SVD threshold:   0.0  
!  NoRefine?:       False  
!                       
! Dispersion coefficients for thiophene (hartree bohr^n)
  C1  C1             C6             C7             C8             C9             C10            C11            C12
    00   00   0   20.73208         0.0          458.1810         0.0          45279.14         0.0          1405874.
  End
  C2  C1             C6             C7             C8             C9             C10            C11            C12
    00   00   0   29.06104         0.0          669.8592         0.0          51070.29         0.0          1633036.
  End
  C2  C2             C6             C7             C8             C9             C10            C11            C12
    00   00   0   41.08062         0.0          998.3487         0.0          55196.48         0.0          1698946.
  End
  S  C1              C6             C7             C8             C9             C10            C11            C12
    00   00   0   45.83737         0.0          1279.625         0.0          61090.16         0.0          2413640.
  End
  S  C2              C6             C7             C8             C9             C10            C11            C12
    00   00   0   64.93124         0.0          1905.714         0.0          61701.34         0.0          1943471.
  End
  S  S               C6             C7             C8             C9             C10            C11            C12
    00   00   0   103.2838         0.0          3563.088         0.0          64177.12         0.0          400303.0
  End
  H1  C1             C6             C7             C8             C9             C10            C11            C12
    00   00   0   6.856521         0.0          130.1041         0.0          5169.989         0.0          95674.96
  End
  H1  C2             C6             C7             C8             C9             C10            C11            C12
    00   00   0   9.694998         0.0          197.4193         0.0          3378.589         0.0          9825.259
  End
  H1  S              C6             C7             C8             C9             C10            C11            C12
    00   00   0   15.50458         0.0          394.6145         0.0         -987.4564         0.0         -280905.2
  End
  H1  H1             C6             C7             C8             C9             C10            C11            C12
    00   00   0   2.345294         0.0          38.21005         0.0         -1376.843         0.0         -44846.70
  End
  H2  C1             C6             C7             C8             C9             C10            C11            C12
    00   00   0   4.946776         0.0          125.4229         0.0          3520.932         0.0          127097.0
  End
  H2  C2             C6             C7             C8             C9             C10            C11            C12
    00   00   0   6.915000         0.0          183.7954         0.0          2170.691         0.0          29222.57
  End
  H2  S              C6             C7             C8             C9             C10            C11            C12
    00   00   0   11.06034         0.0          348.0026         0.0         -607.1553         0.0         -300761.5
  End
  H2  H1             C6             C7             C8             C9             C10            C11            C12
    00   00   0   1.683741         0.0          37.41295         0.0         -1138.014         0.0         -50946.47
  End
  H2  H2             C6             C7             C8             C9             C10            C11            C12
    00   00   0   1.229678         0.0          34.39881         0.0         -838.6934         0.0         -57749.61
  End
\end{lstlisting}
